\documentclass[preprint,5p]{elsarticle}

\usepackage{amssymb}
\usepackage{amsmath}
\usepackage{algorithmic}
\usepackage{algorithm}
\usepackage{multirow}  
\usepackage{graphicx}
\usepackage{booktabs}
\usepackage{colortbl}  
\usepackage{caption}
\usepackage{xcolor} 
\newcommand{\blackfilledcircle}{\raisebox{0.65ex}{\textcolor{black}{\circle*{4}}}}
\usepackage{txfonts}
\usepackage{appendix}

\definecolor{Highlight}{HTML}{39b54a}
\definecolor{shadecolor}{RGB}{150,150,150}
\definecolor{Gray}{gray}{0.5}
\usepackage{adjustbox}

\usepackage[numbers]{natbib}
\usepackage[colorlinks]{hyperref}

\def\tsc#1{\csdef{#1}{\textsc{\lowercase{#1}}\xspace}}
\tsc{WGM}
\tsc{QE}

\begin{document}
\let\WriteBookmarks\relax
\def\floatpagepagefraction{1}
\def\textpagefraction{.001}
\setcounter{topnumber}{3}

\begin{frontmatter}



\title{Innovative Quantitative Analysis for Disease Progression Assessment in Familial Cerebral Cavernous Malformations}


\author[label1,label2]{Ruige Zong\fnref{1}}
\fntext[1]{These authors contributed equally to this work.}
\ead{2277206819@qq.com}

\author[label1,label2]{Tao Wang\fnref{1}}
\ead{ortonwangtao@gmail.com}

\author[label4,label5]{Chunwang Li\fnref{1}}
\ead{lcw@fjmu.edu.cn}

\author[label1,label2,label3]{Xinlin Zhang}

\author[label1,label2]{Yuanbin Chen}

\author[label1,label2]{Longxuan Zhao}

\author[label4,label5]{Qixuan Li}

\author[label1,label2,label3]{Qinquan Gao}

\author[label4,label5,label6,label7,label8]{Dezhi Kang}

\author[label4,label5,label6,label7,label8]{Fuxin Lin\corref{cor1}}
\ead{lfxstudy@126.com}
\cortext[cor1]{Corresponding author}

\author[label1,label2,label3]{Tong Tong\corref{cor2}}
\ead{ttraveltong@gmail.com}
\cortext[cor2]{Corresponding author}

\affiliation[label1]{organization={College of Physics and Information Engineering},
            addressline={Fuzhou University},
            city={Fuzhou},
            state={China}}

\affiliation[label2]{organization={Fujian Key Lab of Medical Instrumentation \& Pharmaceutical Technology},
            addressline={Fuzhou University},
            city={Fuzhou},
            state={China}}
            
\affiliation[label3]{organization={Imperial Vision Technology},
            city={Fuzhou},
            state={China}}

\affiliation[label4]{organization={Department of Neurosurgery, Neurosurgery Research Institute, The First Affiliated Hospital},
            addressline={Fujian Medical University},
            city={Fuzhou},
            state={China}}

\affiliation[label5]{organization={Department of Neurosurgery, National Regional Medical Center, Binhai Campus of the First Affiliated Hospital},
            addressline={Fujian Medical University},
            city={Fuzhou},
            state={China}}
            
\affiliation[label6]{organization={Department of Neurosurgery, Fujian Institute of Brain Disorders and Brain Science, and Fujian Clinical Research Center for Neurological Diseases, First Affiliated Hospital, and Neurosurgery Research Institute},
            addressline={Fujian Medical University},
            city={Fuzhou},
            state={China}}

\affiliation[label7]{organization={Fujian Provincial Clinical Research Center for Neurological Diseases, First Affiliated Hospital},
            addressline={Fujian Medical University},
            city={Fuzhou},
            state={China}}

\affiliation[label8]{organization={Clinical Research and Translation Center, The First Affiliated Hospital},
            addressline={Fujian Medical University},
            city={Fuzhou},
            state={China}}

\begin{abstract}
Familial cerebral cavernous malformation (FCCM) is a hereditary disorder characterized by abnormal vascular structures within the central nervous system. The FCCM lesions are often numerous and intricate, making quantitative analysis of the lesions a labor-intensive task. Consequently, clinicians face challenges in quantitatively assessing the severity of lesions and determining whether lesions have progressed. To alleviate this problem, we propose a quantitative statistical framework for FCCM, comprising an efficient annotation module, an FCCM lesion segmentation module, and an FCCM lesion quantitative statistics module. Our framework demonstrates precise segmentation of the FCCM lesion based on efficient data annotation, achieving a Dice coefficient of 93.22\%. More importantly, we focus on quantitative statistics of lesions, which is combined with image registration to realize the quantitative comparison of lesions between different examinations of patients, and a visualization framework has been established for doctors to comprehensively compare and analyze lesions. The experimental results have demonstrated that our proposed framework not only obtains objective, accurate, and comprehensive quantitative statistical information, which provides a quantitative assessment method for disease progression and drug efficacy study, but also considerably reduces the manual measurement and statistical workload of lesions, assisting clinical decision-making for FCCM and accelerating progress in FCCM clinical research. This highlights the potential of practical application of the framework in FCCM clinical research and clinical decision-making. The codes are available at https://github.com/6zrg/Quantitative-Statistics-of-FCCM.
\end{abstract}




\end{frontmatter}


\section{Introduction}
\label{sec:introduction}
Cerebral cavernous malformation (CCM) is a disease with an abnormal vascular structure in the central nervous system \citep{dalyai2011management, flemming2017population}, characterized by weak vascular walls and slow blood flow, presenting as a "mulberry-like" appearance, surrounded by hemosiderin \citep{zafar2019familial}. CCM can be categorized into two types based on the number of lesions: the first is sporadic, typically presenting as a single lesion; the second is multifocal, characterized by multiple lesions within the brain. The majority of multifocal CCM results from autosomal dominant genetic mutations, indicating a hereditary nature. This type of CCM is referred to as familial cerebral cavernous malformation (FCCM) \citep{incerti2023magnetic}. According to literature reports, the prevalence of CCM ranges from approximately 0.100\% to 0.800\% \citep{haasdijk2012cerebral, mouchtouris2015management}. The estimated population prevalence of FCCM is between 0.026\% and 0.030\% \citep{spiegler2018cerebral}.

FCCM frequently results in cerebral hemorrhage, seizures, and other neurological functional impairments \citep{weinsheimer2023intracranial}. Nevertheless, the mechanisms triggering these symptoms remain unclear, and the severity of the disease can vary greatly even among individuals within the same family or similar age. Moreover, there is currently a lack of a well-established therapeutic approach for FCCM. The hereditary and refractory nature of FCCM imposes significant economic and psychological burdens on society and families.

Currently, magnetic resonance imaging (MRI) stands as the optimal method for assessing CCM. The susceptibility-weighted imaging (SWI) sequences in MRI demonstrate the highest sensitivity to this condition, enabling the detection of minute lesions not discernible in T1 and T2 sequences. This capability is valuable for estimating the quantity and volume of FCCM lesions, thereby facilitating the assessment of disease burden, studies on genetic mutation phenotypes, and providing a basis for clinical decision-making and treatment \citep{zafar2019familial}. 

\begin{figure}
	\centering
	\includegraphics[width=0.478\textwidth, angle=0]{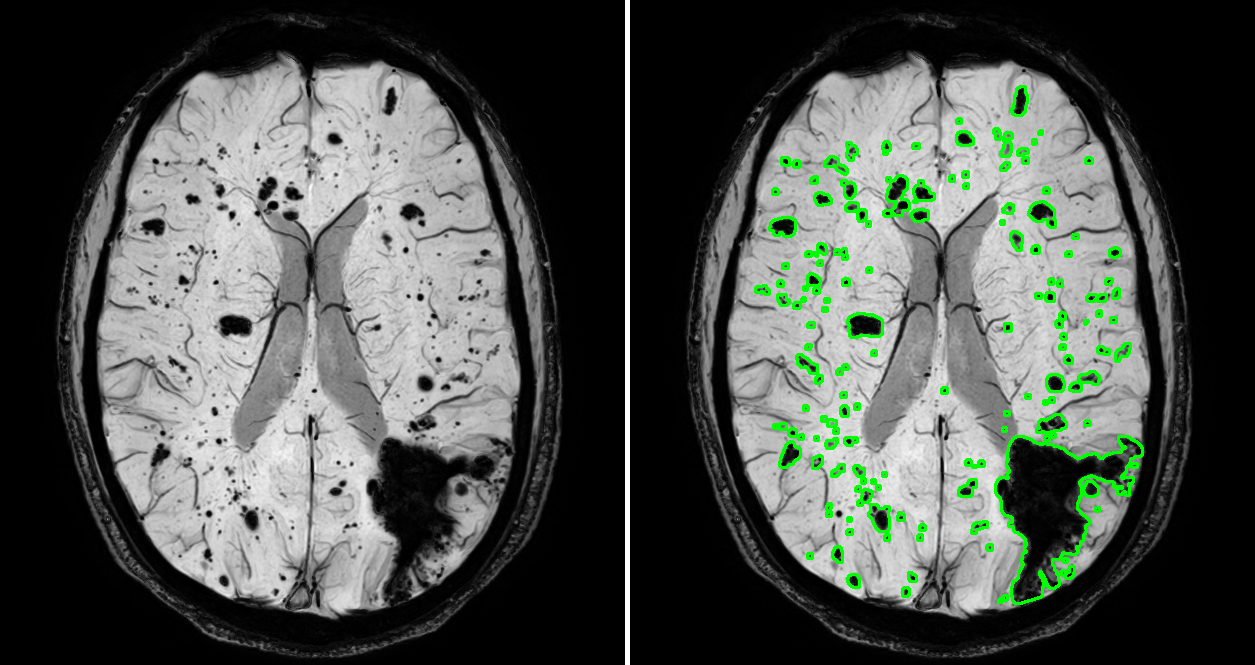}
	  \caption{Complex FCCM lesion cases. The green markings represent the regions of lesions.}
    \label{1}
    \vspace{-0.5em}
\end{figure}

However, presently, statistical analysis of lesion volume and quantity in FCCM relies predominantly on manual measurement methods. Due to the complexity and large number of lesions, especially the presence of numerous small lesions as illustrated in Figure \ref{1}, manual measurement of their volume and quantity is almost impossible. Consequently, current manual measurements primarily focus on larger lesions, yet are unavoidably subject to the influence of physicians' subjective cognition, resulting in certain measurement inaccuracies. Moreover, due to the large number of lesions, clinicians find it challenging to detect progressive lesions with the naked eye. Therefore, there is an urgent demand for an objective and precise method to quantitatively assess lesion volume and quantity in FCCM patients. Simultaneously, it is anticipated that a reference for identifying clearly progressing lesions between two MRI images can be provided to clinicians by this method.

In recent years, the outstanding performance of deep learning algorithms has made them the preferred choice for medical image segmentation. The ongoing advancements in medical imaging technology further enhance the applicability of deep learning in the medical domain. The emergence of end-to-end fully convolutional networks \cite{long2015fully} has significantly propelled the advancement of medical image segmentation. Ronneberger et al. \cite{ronneberger2015u} proposed the U-Net framework, which has become popular neural network architecture for biomedical image segmentation tasks \citep{norman2018use, sevastopolsky2017optic, roy2017relaynet, skourt2018lung}. Then, an increasing number of researchers have delved deeper into the U-Net architecture, yielding a succession of novel research outcomes \citep{zhang2018road, apostolopoulos2017pathological, alom2018recurrent, hoorali2022irunet}. However, the application of deep learning in the medical field still faces numerous challenges, such as sample scarcity, difficulty in data annotation, and imbalanced sample features. These all increase the difficulty of applying deep learning to medical images.

FCCM is a rare disease. In this study, we collect 101 volumes of FCCM based on MRI SWI sequence, representing the largest sample size in current deep learning research on FCCM. Based on this, a rapid and efficient annotation method tailored to this dataset is proposed, concurrently leveraging segmentation predictions to further enhance our annotation efficiency. The annotated results undergo quality assessment by clinicians to ensure the accuracy and reliability of our data annotations. Furthermore, we introduce a deep learning framework for image segmentation tailored to FCCM and propose a quantitative statistics module of FCCM lesions, which realizes the quantitative statistics of lesion segmentation results, encompassing the number of lesions and the volume of each lesion.

Furthermore, combining image registration, we perform a statistical assessment of lesions between two examinations for the same patient. This aids physicians in gaining a better understanding of the patient's disease progression. Additionally, the module includes the quantitative analysis visualization of lesions, with each lesion annotated in the visualization alongside its corresponding volume information, facilitating clinicians in promptly discerning progressing lesions.

In summary, the main contributions of this paper are as follows:

\blackfilledcircle{} A rapid data annotation method designed for FCCM is proposed, which is convenient and efficient compared to traditional manual contour tracing labeling. Physician assessments ensure the reliability of our data annotation.

\blackfilledcircle{} We employ deep learning to achieve precise segmentation of FCCM lesions and conduct a comparative evaluation against widely used medical image segmentation methods for FCCM lesion segmentation.

\blackfilledcircle{} A quantitative statistical method for the volume and quantity of FCCM lesions is implemented, and we conduct comparative statistics on lesions from two examinations of the same patient employing image registration. To our knowledge, our work represents a pioneering effort in employing artificial intelligence for the quantitative statistics of FCCM. This provides a precise quantitative assessment basis for disease progression and drug efficacy.

\blackfilledcircle{} We establish a visualization framework to comprehensively showcase the segmentation and quantitative statistical results of FCCM lesions, offering physicians an intuitive analytical basis for the clinical diagnosis and research of FCCM.

\begin{figure*}[h]
	\centering
	\includegraphics[scale=0.08]{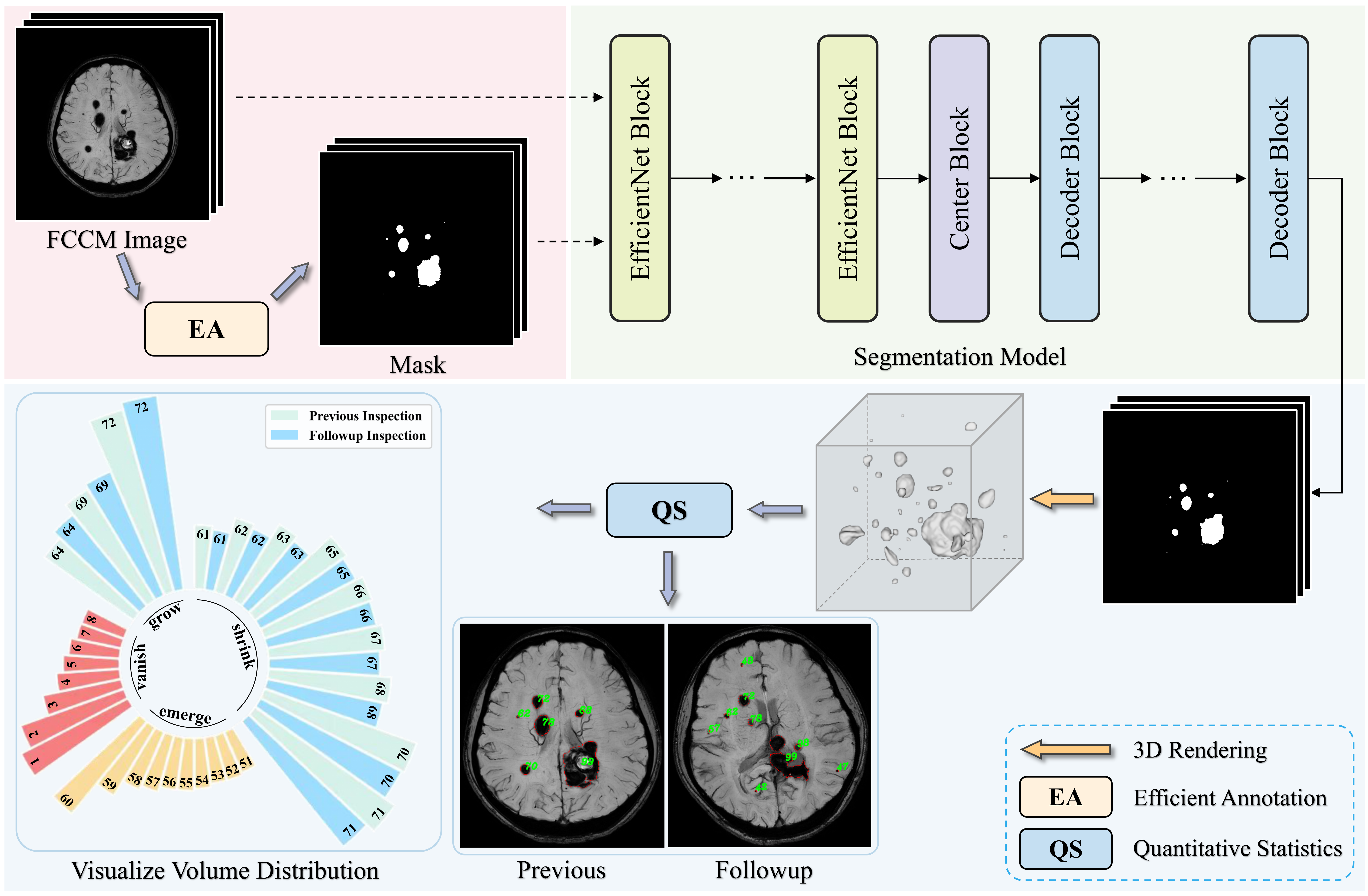}
	\caption{FCCM quantitative statistical framework. The visualization of volume distribution (see Figure \ref{9} in Appendix A for details) displays the indexes and volume information of lesions. The "grow" and "shrink" categories respectively represent the volume pairs of lesions with increased or decreased volume between two inspections. The "emerge" and "vanish" categories depict the lesions that are newly developed or disappeared (or unsuccessfully matched) compared to the previous inspection.}
	\label{2}
    \vspace{-0.5em}
\end{figure*}

\section{Related work}
\subsection {FCCM research}
As a noteworthy neurosurgical disorder, FCCM has garnered extensive attention in recent research. Flemming et al. \citep{flemming2020cerebral} have emphasized crucial information that clinicians should comprehend, encompassing clinical presentations, imaging characteristics, genetic factors, and treatment options related to FCCM. In addition, numerous studies have also demonstrated the importance of quantitative statistics in FCCM. Fox et al. \citep{fox2021seizure} investigated the correlation between CCM counts or genotypes and early-onset seizures, demonstrating that individuals with a higher number of CCM lesions or the CCM3 genotype have a higher incidence of seizures. Cheng et al. \citep{cheng2021fetal} underscored that the treatment approach for FCCM patients should be determined based on the severity of the condition quantitatively assessed and the location of the lesions.

Weinsheimer et al. \citep{weinsheimer2023intracranial} pointed out that the current counting methods for FCCM lesions mostly manually calculate the number of larger lesions (maximum diameter $\geq$ 5 mm) in T2 sequences. Incerti et al. \citep{incerti2023magnetic} employed 3T MRI scans for the semi-automatic segmentation of FCCM and calculated the maximum quantitative susceptibility mapping for each segmented CCM in the baseline. However, the acquisition of this quantitative metric involves intricate specialized processing and has not yet become commonplace in clinical practice. Therefore, an objective and accurate quantitative statistical method is proposed for lesion volume and quantity based on the segmentation results for FCCM lesions.

\begin{figure*}[h]
	\centering
	\includegraphics[scale=0.529]{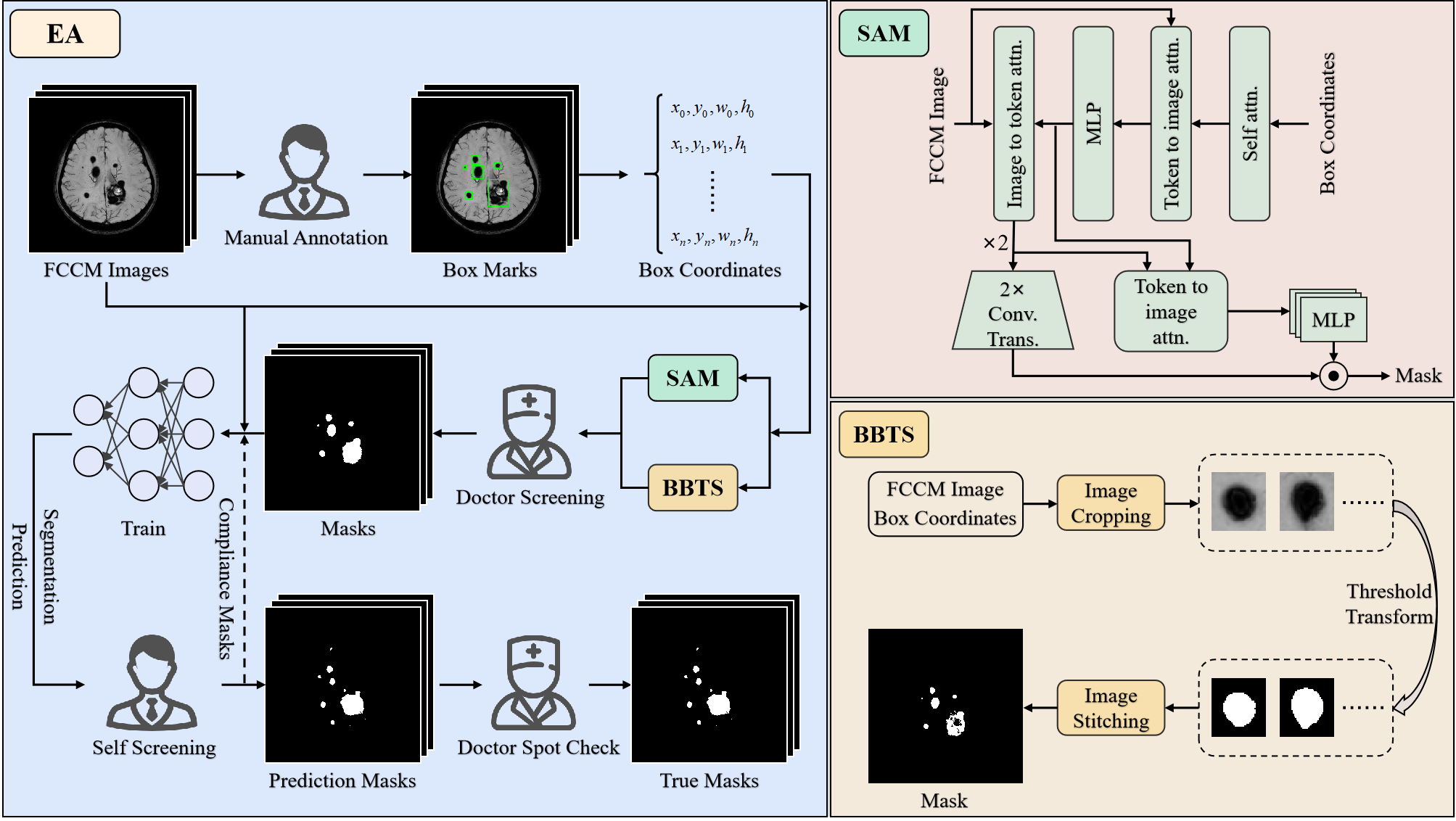}
	  \caption{Efficient annotation module.}
    \label{3}
    \vspace{-0.5em}
\end{figure*}

\subsection{Annotation studies in segmentation}
In recent years, medical image segmentation plays an important role in the field of modern medicine and gradually develops towards the direction of actual medical diagnosis, providing support for diagnosis and treatment decision-making. For example, Lan et al. \citep{lan2023using} established a deep learning-based pathological auxiliary diagnostic system for gastric cancer, enhancing diagnostic accuracy and efficiency among pathologists.

Due to the highly intricate nature of medical images, annotating medical data is labor-intensive and expensive. Consequently, the application of image segmentation in medical research is constrained. In order to mitigate annotation costs, many researchers have proposed a series of segmentation-assisted annotation strategies \citep{elaraby2021framework, philbrick2019ril, deng2019new, kline2016semiautomated, racimora2010segmentation, yushkevich2006user}. For instance, Deng et al. \citep{deng2019new} introduced a novel medical image annotation framework that integrates active learning and U-shaped networks, obtaining the most effective annotation candidates through iterative training. Kline et al. \citep{kline2016semiautomated} employed a loosely defined delineation of renal boundaries as a cue, achieving semi-automated segmentation annotation. Racimora et al. \citep{racimora2010segmentation} investigated a semi-automated segmentation annotation method based on active contours and sub-voxel morphological algorithms. The segmentation was conducted through the application of the Bridge Burner algorithm to determine morphological thresholds. Furthermore, Kirillov et al. \citep{sam2023segment} proposed the Segment Anything (SAM) segmentation model, which achieves target region segmentation based on points, bounding boxes, text, and masks as prompts and is the first fundamental large model dedicated to image segmentation. By undergoing pre-training on the ultra-large-scale SA-1B dataset containing over 11 million images and 1 billion masks \citep{sam2023segment}, SAM demonstrates outstanding zero-shot transfer capabilities.

We use a bounding box that completely covers the target as a prompt for semi-automatic segmentation to annotate partial data. Subsequently, the remaining data is annotated through iterative training of a segmentation network. Threshold segmentation is employed as one of the techniques for annotation based on bounding box prompts, while the SAM model segmentation serves as another method.

\section{Methods}
This section introduces the MRI-based SWI sequence dataset (SG-FCCM) and the quantitative statistical framework for FCCM. The framework is primarily composed of three components: the efficient annotation module (EA), the FCCM lesion segmentation module, and the FCCM lesion quantitative statistical module (QS), as illustrated in Figure \ref{2}. Given the complexity and time-consuming nature of medical image annotation, an efficient annotation method is proposed to annotate FCCM images. Subsequently, the annotated data is utilized for training the FCCM segmentation model. In the QS module, the pre-trained segmentation network is employed for FCCM lesion segmentation, quantifying the volume and quantity of lesions through the segmentation masks. Furthermore, to understand the disease progression of patients, contrastive statistics of lesions across different examinations for a patient are achieved in conjunction with image registration. Finally, the segmentation results and corresponding volume information of the lesion are displayed in the quantitative analysis visualization, facilitating diagnostic analysis for clinicians. Specific details will be expounded upon in the following sections.

\subsection{Dataset}
We collect a total of 101 volumes of FCCM from 73 patients, constituting the SG-FCCM dataset. Among these, 26 patients have data for both enrollment and a 1-year follow-up. These data are 3D image data generated based on the SWI sequence of MRI. The 3D images are decomposed into 2D slices, resulting in 4089 2D images with dimensions of 768 $\times$ 768 pixels, on which annotations are conducted. Finally, random partitioning allocates approximately 10\% to the validation set (400 images), 10\% to the test set (400 images), and the remaining 80\% to the training set (3289 images). The SWI parameters are as follows: time repetition (TR) 31 ms, echo time (TE) 7.2 ms, deltaTE 6.2 ms, field of view (FOV) 200 $\times$ 230 mm$^{2}$, matrix 384 $\times$ 332, slices 130, and slices thickness 2 mm.

\subsection{Efficient annotation module}
The EA module architecture and the implementation of annotation methods are shown in Figure \ref{3}. The data annotation process is shown in Algorithm \ref{alg:Data_Annotation}. Primarily, we annotate lesions in 1579 images of 39 volumes using bounding boxes, employing the SAM model and box binarization threshold segmentation (BBTS) for segmentation annotation. Following physician screening, 1378 qualified masks are obtained and used to train a segmentation network. Secondly, the remaining 201 images with unqualified annotations are iteratively trained alongside 2510 images without annotated bounding boxes. It is worth noting that during each iteration, partial segmentation masks are screened, and qualified masks are merged into the training data. Subsequently, iterative training is performed to optimize the segmentation performance of the network, thereby obtaining more qualified masks. In addition, after multiple iterations, 561 images remain challenging to obtain ideal masks directly from the segmentation network. Therefore, this portion of the mask is subjected to semi-automatic annotation correction by experienced doctors, including deleting over-segmented areas and filling up under-segmented areas.

\begin{algorithm}[!h]
    \caption{Data Annotation Process}
    \label{alg:Data_Annotation}
    \renewcommand{\algorithmicrequire}{\textbf{Input:}}
    \renewcommand{\algorithmicensure}{\textbf{Output:}}
    \begin{algorithmic}[1]
    \REQUIRE FCCM images, $X = \left \{ x_{1}, x_{2}, ..., x_{4089} \right \}$
    \ENSURE FCCM masks, $Y = \left \{ y_{1}, y_{2}, ..., y_{4089} \right \}$
    \vspace{0.3\baselineskip}
    \FOR{$x_{i}$ in $\left \{ x_{1}, x_{2}, ..., x_{1579} \right \}$ }
        \STATE $b_{i}$ = Annotated bounding boxes of the lesions in $x_{i}$
        \STATE $m_{i}$ = SAM $(x_{i}, b_{i})$
        \STATE $n_{i}$ = BBTS $(x_{i}, b_{i})$
        \STATE $k_{i}$ = Doctor Screening $(x_{i}, m_{i}, n_{i})$
    \ENDFOR

    \STATE \textcolor{blue}{\# $t$ represents the number of iterative training}
    \STATE \textcolor{blue}{\# $X_{j}$ is the collection of FCCM images corresponding to the qualified masks in set $K_{j}$}
    \FOR{$j = 0$ to $t$}
    \IF{$j = 0$}
        \STATE $K_{j}$ = $\left \{ k_{1}, k_{2}, ..., k_{1378} \right \}$
    \ELSE
        \STATE $K_{j}$ = $Q_{j-1}$ add to $K_{j-1}$
    \ENDIF
    \STATE $N_{seg}$ = Train $(X_{j}, K_{j})$
    \STATE $P_{j}$ = Predict $(N_{seg}, X - X_{j})$
    \STATE $Q_{j}$ = Self Screening $(P_{j})$
    \IF{$j = t$}
        \STATE $K_{t}$ = $Q_{t}$ add to $K_{t}$
    \ENDIF
    \ENDFOR
 
    \STATE After iterative training, there are 561 unsatisfactory masks remaining, $W = \left \{ w_{1}, w_{2}, ..., w_{561} \right \}$
    \STATE $Z$ = Doctor Correct $(W)$
    \STATE $Y$ = $K_{t} + Z$
    \RETURN FCCM masks, $Y = \left \{ y_{1}, y_{2}, ..., y_{4089} \right \}$
    \end{algorithmic}
\end{algorithm}

\subsubsection{SAM model annotation}
The annotation with the SAM model is based on the prompt encoder framework provided by the large model, which allows the segmentation of target regions using bounding box prompts. We acquire coordinate data corresponding to annotated bounding boxes, specifying the coordinates of the top-left and bottom-right vertices. Subsequently, the box-guided target segmentation method of the SAM model is employed to segment the lesion regions, thereby achieving fast and efficient segmentation annotation.

\subsubsection{BBTS annotation}
In the collected FCCM data, there is a distinct color contrast between lesions and the background. Therefore, a binarization threshold segmentation method is utilized for FCCM lesions. Initially, binarization segmentation is applied to the entire image, but the determination of the threshold is challenging, resulting in inaccurate segmentation outcomes. Therefore, the box binarization threshold segmentation is adopted. This method restricts the binarized region to a bounding box, making it easier to select an appropriate threshold. Additionally, efforts are made during the annotation of the bounding box to ensure significant pixel value differences between the lesion and non-lesion areas within the bounding box, aiming to achieve better segmentation outcomes.

For BBTS annotations, the coordinates of marked bounding boxes are used to extract corresponding patches from the original FCCM image. Subsequently, we apply binarization threshold segmentation to these patches, and the segmentation results of each patch are superimposed onto the background image with the same dimensions as the original FCCM image, completing the segmentation annotation.

\begin{figure*}[h]
	\centering
	\includegraphics[scale=0.559]{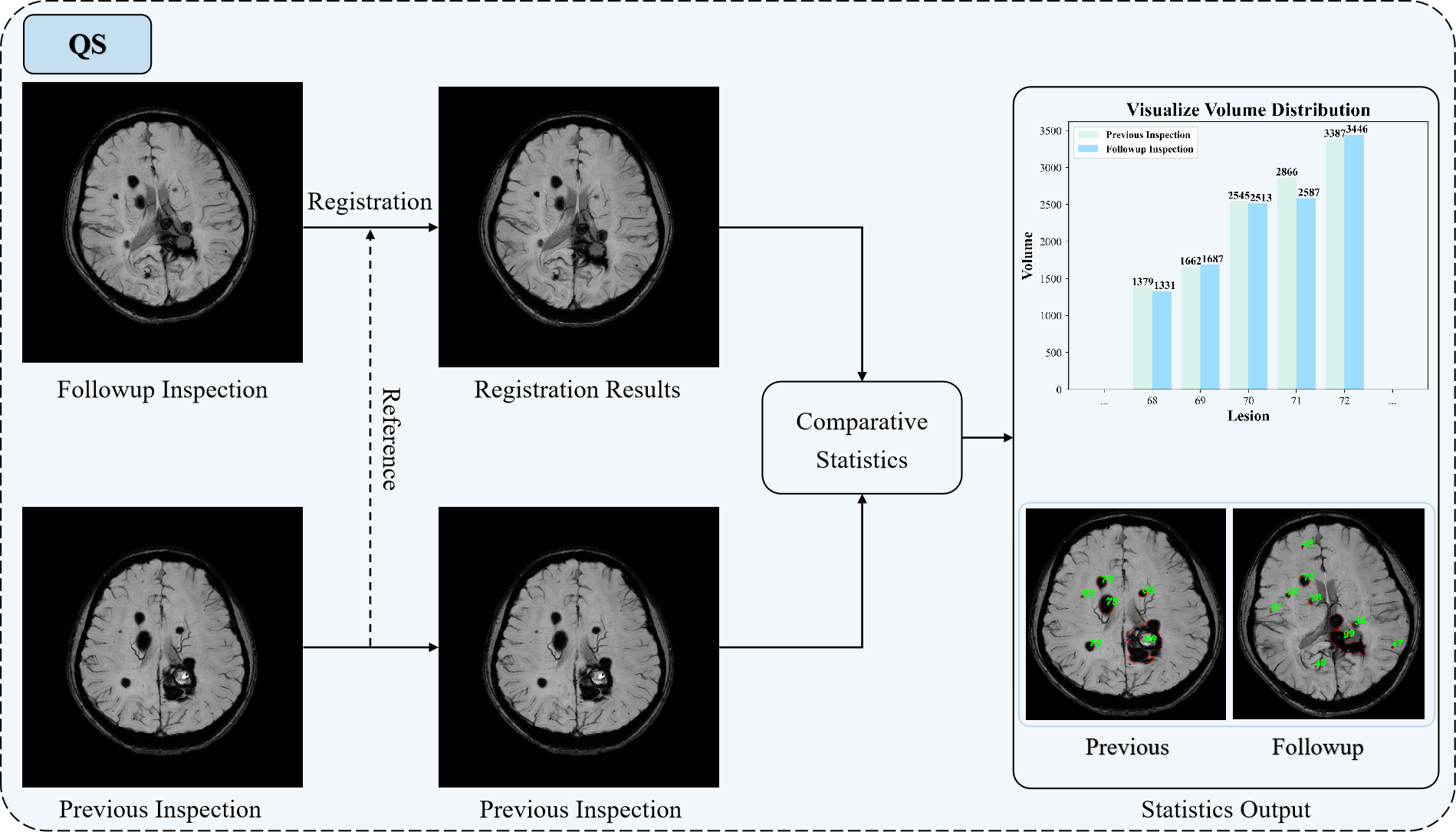}
	  \caption{FCCM lesion quantitative statistical module.}
    \label{4}
    \vspace{-0.5em}
\end{figure*}

\subsection{FCCM lesion segmentation network}
To quickly identify and locate the FCCM lesion areas for quantitative statistics of lesions, a neural network suitable for FCCM lesion segmentation is introduced. As shown in Figure \ref{2}, in the segmentation of FCCM lesions, the U-Net \cite{ronneberger2015u} architecture is utilized as the backbone, with EfficientNet-B3 \cite{tan2019efficientnet} serving as the encoder. This encoder utilizes mobile reverse bottleneck convolution, composed of deep separable convolutional layers. It comprises four convolutional layers and an ordinary convolutional layer with a 2 $\times$ 2 stride, resulting in a halving of the output feature map size at each stage. Intermediate feature maps from different stages of the encoder are extracted for skip connections. The image is input into the network at its original size to minimize information loss. The U-Net structure, encompassing up-sampling and convolution operations, ensures the output image dimensions align with those of the input image.

\subsection{FCCM lesion quantitative statistical module}
The manual measurement and statistical methods for FCCM lesions are not only tedious and time-consuming. To achieve fast and accurate quantitative statistics of FCCM lesions and provide an intuitive diagnostic analysis basis for the condition, the FCCM lesion quantitative statistics module is proposed, as illustrated in Figure \ref{4}, which comprises three components: image registration, lesion quantitative comparative statistics, and quantitative analysis visualization. Algorithm \ref{alg:Quantitative_Statistical} outlines the quantitative statistical process for lesions. Lesion segmentation is performed on slices from two examinations, with 2D masks being synthesized into a 3D mask. Subsequently, the volume and quantity of lesions are quantified, and the coordinates of the minimum bounding cube for each lesion are stored. Combining image registration, lesions from the follow-up inspection are sequentially matched with those from the previous inspection, completing comparative statistics. Ultimately, a visualization of the segmentation results and statistical information is generated.

\begin{algorithm}[!h]
    \caption{Quantitative Statistical Process}
    \label{alg:Quantitative_Statistical}
    \renewcommand{\algorithmicrequire}{\textbf{Input:}}
    \renewcommand{\algorithmicensure}{\textbf{Output:}}
    \begin{algorithmic}[1]
    \REQUIRE $X_{p}$ (Previous inspection), $X_{f}$ (Followup inspection)
    \ENSURE Volume statistical information, Quantitative analysis visualization
    \vspace{0.3\baselineskip}
    \STATE $U_{p}$ = Lesions statistics $(X_{p})$, $U_{p}$ = $\left \{ u_{p1}, u_{p2}, ..., u_{pm} \right \}$
    \STATE $U_{f}$ = Lesions statistics $(X_{f})$, $U_{f}$ = $\left \{ u_{f1}, u_{f2}, ..., u_{fn} \right \}$
    \STATE \textcolor{blue}{\# $m$, $n$ represents the number of lesions}
    \STATE \textcolor{blue}{\# $u$ = [$V$, $B$], $V$ is lesion volume, $B$ is minimum bounding cube coordinates}
    \FOR{$u_{fi}$ in $\left \{ u_{f1}, u_{f2}, ..., u_{fn} \right \}$ }
        \STATE $z$ = Get lesion center coordinates $(u_{fi}[B])$
        \STATE $X^{'}_{f}$ = Set specific pixel $(z, X_{f})$
        \STATE $X^{*}_{f}$ = Register $(X_{p}, X^{'}_{f})$
        \STATE $z^{*}$ = Get specific pixel coordinates $(X^{*}_{f})$
        \STATE $U^{*}_{f}$ = Lesions statistics $(X^{*}_{f})$
        \STATE \textcolor{blue}{\# Matching $u_{fi}$ in $X_{f}$ and $X^{*}_{f}$ by $z^{*}$ and in $X_{p}$ and $X^{*}_{f}$ by the comparison between $U_{p}$ and $U^{*}_{f}$ . Matching successfully outputs volume pair, otherwise only outputs $u_{fi}[V]$}
        \STATE $\{(u_{pi}[V], u_{fi}[V])/ u_{fi}[V]\}$ = Comparative statistics $(U^{*}_{f}, U_{p}, U_{f}, z^{*})$
    \ENDFOR
    \STATE Visualization generation $(X_{f}$ 2D slices, 2D prediction masks, $\{(U_{p}[V], U_{f}[V])/ U_{f}[V]\})$
    \RETURN $\{(U_{p}[V], U_{f}[V])/ U_{f}[V]\}$, Quantitative analysis visualization
    \end{algorithmic}
\end{algorithm}

\begin{table*}[]
\caption{\label{arttype}Summary of visualization results.}
\centering
\tabcolsep=1.55cm
\begin{tabular}{|c|c|}
\hline
Visualization results & Index \\
\hline
visualization collages & Figure \ref{7}, Figure \ref{10} and Figure \ref{11} in the Appendix A \\
\hline
quantitative statistical table & Table \ref{Table2} \\
\hline
disease assessment line chart & Figure \ref{8} \\
\hline
volume division statistics table & Table \ref{Table3} \\
\hline
\end{tabular}
\label{Table}
\vspace{-0.5em}
\end{table*}

\subsubsection{Image registration} Due to potential discrepancies in the positioning of head scans between two examinations, the same lesion may exhibit positional variations in the respective images, posing challenges for matching each other. Therefore, image registration is employed to address this problem. Considering the potential information loss associated with registration, the patient's first examination image is used as the reference, aligning the second examination image to it, rather than concurrently registering both images to a standard brain template, thereby mitigating errors. Additionally, special pixel markers are incorporated before and after registration to overcome potential minor variations in lesion volume resulting from the registration process, thus providing a true volume comparison.

\subsubsection{Lesion quantitative comparative statistics} To compare variations in lesion volume at different inspection stages and assist physicians in gaining a better understanding of the patient's disease progression, comparative statistics of lesions from different inspections are performed. Primarily, the calculation of lesion quantity and pixel number (volume) is performed by detecting positive segmentation regions (lesion areas) within 3D masks. The volumes of lesions from previous and follow-up inspections are documented, and the coordinates of the minimum bounding cube for each lesion are concurrently acquired and stored. Inspired by the concept of Intersection over Union (IoU) \citep{10.1007/978-3-319-50835-1_22, Rezatofighi_2019_CVPR}, we propose Intersection over Boundary Cube (IoC), which is the intersection volume between two bounding cubes divided by the volume of the larger bounding cube. IoC can be formulated as follows:
\begin{align}
    &IoC = \frac{V_{intersection}}{V_{larger}}
\end{align}
where $V_{intersection}$ and $V_{larger}$ represent the intersection volume of two lesion bounding cubes and the volume of the larger bounding cube, respectively.

The lesions in the follow-up inspection are traversed, and the IoC values are calculated between the traversed lesion and each of the lesions in the previous inspection. Given that image registration achieved alignment of brain images from the two examinations, ensuring accurate spatial correspondence of the lesions. Furthermore, the introduction of IoC considers both the overall morphology and position of the lesions, and the intersection of the minimum bounding cubes indicates relatively minor changes in the position and morphology of the lesion area between the two examinations. This consistency enhances the accuracy of matching. Therefore, we consider two lesions with IoC values greater than 0 to be matched as the same lesion region. In case the IoC values of a traversed lesion and multiple lesions in the previous examination are all greater than 0, the lesion pair with the maximum IoC value is matched as the same lesion region, and the changes in lesion volume are presented in the form of volume pairs. If during a traversal, no IoC value is greater than 0, it is considered that this lesion in a follow-up inspection does not successfully match any lesion from the previous inspection. In such cases, the volume of this lesion in the follow-up examination is presented separately.

\subsubsection{Quantitative analysis visualization} To facilitate lesion observation by clinicians, the contours of the lesions are delineated on the original FCCM images using the segmentation masks. Lesions are sequentially numbered, and visualization collages for the two examinations are generated. Additionally, we create a quantitative statistical table for lesion volumes, plot an overall disease assessment line chart based on the lesion volume information between different examinations, and calculate the proportion of lesion volume in different ranges. This ultimately forms a visualization framework to assist doctors in comprehensively analyzing disease progression. This framework enables physicians to quickly locate lesion areas and perform faster and more accurate quantitative assessments of lesion changes between different examinations, thus better understanding the progression of the patient's condition. All the visualization results provided to physicians are presented in Table \ref{Table}.

\section{Experiments and results}
\subsection{Data annotation results}
\begin{figure}
	\flushleft
	\includegraphics[width=0.463\textwidth, angle=0]{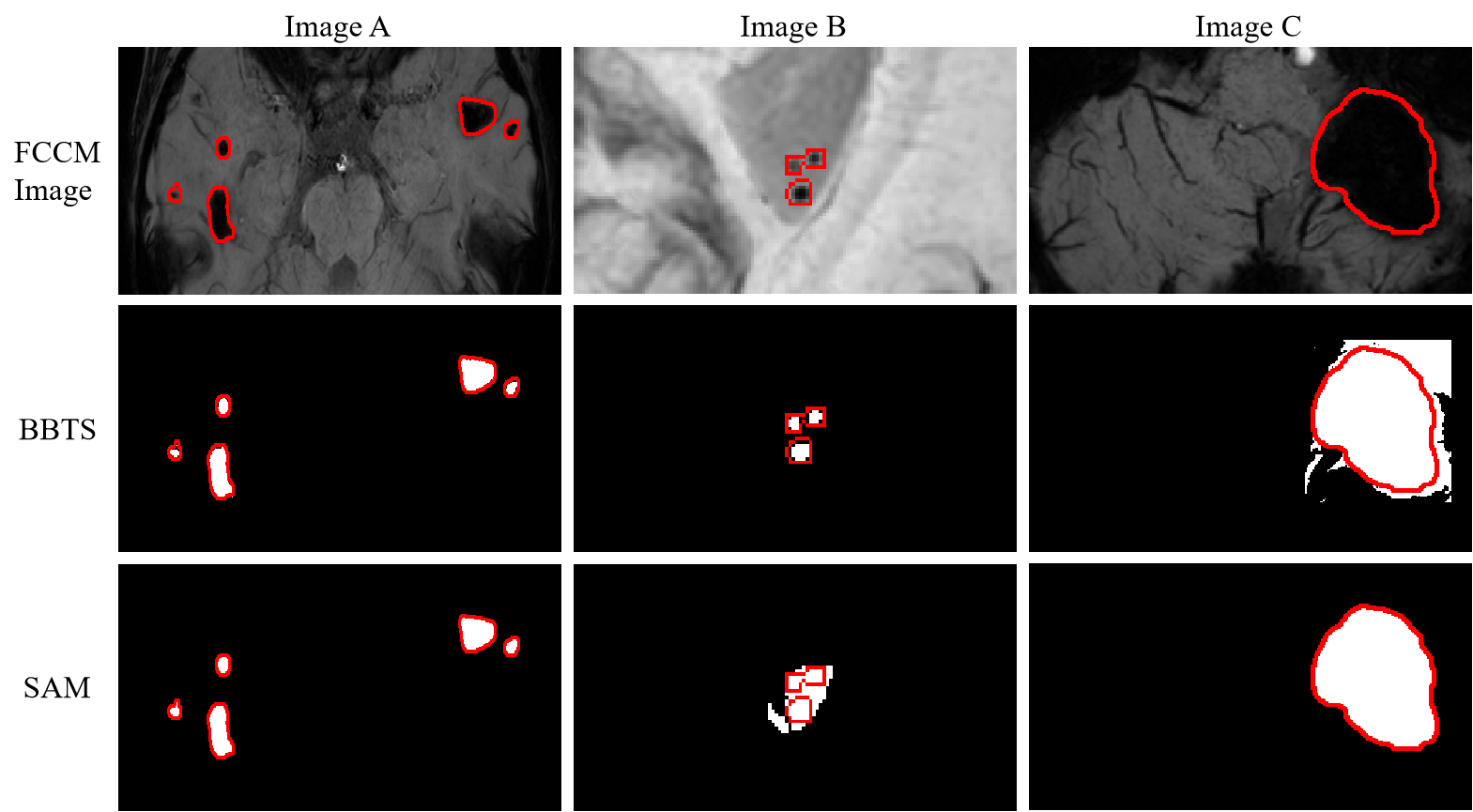}
	  \caption{BBTS/SAM annotation result comparison.}
    \label{5}
    \vspace{-0.5em}
\end{figure}

In the EA module, we initially annotated the lesions in 1579 images with bounding boxes. Subsequently, employing two segmentation methods resulted in a total of 1378 qualified masks. Among them, 191 were only qualified under BBTS annotation, and 185 were only qualified under SAM annotation. Figure \ref{5} illustrates the segmentation annotation results of the two methods. In which, image A was satisfactorily annotated under both segmentation methods, and images B and C respectively showed examples of qualified annotations only under BBTS and SAM annotations. The mask qualification rate for box prompt segmentation annotation reached 87.3\%. Then a total of 2711 images underwent iterative training annotation and the 2150 qualified candidate masks and 561 unqualified masks were screened. The candidate masks were subjected to further quality assessment by doctors, resulting in 2117 qualified masks. The final iterative training labeling qualification rate reached 78.1\%.

\begin{table*}[]
\caption{\label{arttype1}Segmentation performance of different methods on the SG-FCCM dataset.}
\centering
\resizebox{\textwidth}{!}{%
\begin{tabular}{cccccccccc}
\hline
Method & Encoder & Dice(\%) & mIoU(\%) & Precision(\%) & Recall(\%) & F1-Score(\%) & FLOPs(G) & Parameters(M) \\
\hline
U-Net \cite{ronneberger2015u} & \multirow{6}{*}{-} & 91.82 & 89.33 & 89.22 & 87.26 & 83.66 & 723.43 & 17.26 \\
CE-Net \cite{gu2019net} & & 85.14 & 82.41 & 79.96 & 80.31 & 70.32 & 160.55 & 29.00 \\
Swin-Unet \cite{cao2022swin} & & 90.10 & 87.51 & 86.81 & 85.28 & 80.22 & 139.15 & 27.15 \\
TransFuse \cite{zhang2021transfuse} & & 88.82 & 86.15 & 86.04 & 83.27 & 77.66 & 207.59 & 26.17 \\
TransUNet \cite{chen2021transunet} & & 89.33 & 86.83 & 86.28 & 84.73 & 78.70 & 584.36 & 93.23 \\
UNeXt \cite{valanarasu2022unext} & & 88.11 & 85.44 & 83.32 & 84.75 & 76.24 & \textbf{10.33} & 1.47 \\
\hline
\multirow{3}{*}{DeepLabv3+ \cite{chen2018encoder}} & ResNet34 \cite{he2016deep} & 90.36 & 87.63 & 86.69 & 85.78 & 80.75 & 142.68 & 22.44 \\
& MobileNetV2 \cite{howard2017mobilenets} & 89.88 & 87.23 & 86.63 & 84.88 & 79.78 & 27.78 & 4.38 \\
& EfficientNet-B3 \cite{tan2019efficientnet} & 92.04 & 89.38 & 89.55 & 86.10 & 84.10 & 11.68 & \textbf{1.07} \\
\hline
\multirow{3}{*}{UNet++ \cite{zhou2018unet++}} & ResNet34 \cite{he2016deep} & 92.87 & 90.53 & 91.05 & 88.49 & 85.77 & 332.27 & 26.08 \\
& MobileNetV2 \cite{howard2017mobilenets} & 92.20 & 89.80 & 90.11 & 87.80 & 84.42 & 81.19 & 6.82 \\
& EfficientNet-B3 \cite{tan2019efficientnet} & 92.91 & 90.54 & 90.11 & \textbf{88.93} & 85.84 & 118.42 & 3.01 \\
\hline
\multirow{3}{*}{U-Net \cite{ronneberger2015u}} & ResNet34 \cite{he2016deep} & 91.30 & 88.85 & 87.81 & 87.78 & 82.62 & 141.47 & 21.44 \\
& MobileNetV2 \cite{howard2017mobilenets} & 92.57 & 90.18 & 91.14 & 87.24 & 85.16 & 61.30 & 6.63 \\
& \cellcolor{lightgray!40}\textbf{EfficientNet-B3 \cite{tan2019efficientnet}} & \cellcolor{lightgray!40}\textbf{93.22} & \cellcolor{lightgray!40}\textbf{90.76} & \cellcolor{lightgray!40}\textbf{92.19} & \cellcolor{lightgray!40}87.61 & \cellcolor{lightgray!40}\textbf{86.45} & \cellcolor{lightgray!40}48.48 & \cellcolor{lightgray!40}2.55 \\
\hline
\multicolumn{1}{l}{} & \multicolumn{1}{l}{} & \multicolumn{1}{l}{} & \multicolumn{1}{l}{} & \multicolumn{1}{l}{} & \multicolumn{1}{l}{}
\end{tabular}
}
\label{Table1}
\vspace{-1em}
\end{table*}

\subsection{Experiments of FCCM lesion segmentation}
\subsubsection{Evaluation metrics}
To evaluate segmentation performance, the Dice coefficient (Dice), mean Intersection over Union (mIoU), Precision, Recall, and F1-score are used as evaluation metrics. The metrics are calculated as follows:
\begin{align}
    &Dice = \frac{2 \times TP}{2 \times TP+FN+FP} \label{eq:dice} \\
    &mIoU = \frac{TP}{TP+FN+FP} \\
    &Precision = \frac{TP}{TP+FP} \\
    &Recall = \frac{TP}{TP+FN} \\
    &F1-score = 2 \times \frac{Recall \times Precision}{Precision+Recall}
\end{align}
where TP, TN, FP, and FN are true positives, true negatives, false positives, and false negatives, respectively.

\subsubsection{Implementation details}
The segmentation network was developed in the environment of Python 3.8 and PyTorch 1.12.1. The training and testing platform utilized an Ubuntu 20.04 system with an Nvidia RTX 3070 GPU, featuring a memory capacity of 8 GB. The SG-FCCM dataset was employed for training the segmentation network, with data augmentation techniques such as flipping and rotation to enhance the model's generalization capability. The input image size was set to 768 $\times$ 768, preserving the original image size to minimize information loss. Parameters of the EfficientNet blocks were initialized using weights pretrained on ImageNet \cite{deng2009imagenet}. During the training phase, we set the initialized learning rate to 0.0001, the batch size to 2, and the maximum epoch number to 50 and the AdamW optimizer was employed for model optimization.

\subsubsection{Experimental results}
In segmentation experiments, ResNet34 \cite{he2016deep}, EfficientNet-B3 \cite{tan2019efficientnet}, and MobileNetV2 \cite{howard2017mobilenets} were respectively employed as encoders. U-Net \cite{ronneberger2015u}, UNet++ \cite{zhou2018unet++}, and DeepLabv3+ \cite{chen2018encoder} were used as decoders to construct segmentation networks to select the optimal model in FCCM lesion segmentation. Additionally, segmentation models, including U-Net \cite{ronneberger2015u}, CE-Net \cite{gu2019net}, Swin-Unet \cite{cao2022swin}, TransFuse \cite{zhang2021transfuse}, TransUNet \cite{chen2021transunet}, and UNeXt \cite{valanarasu2022unext}, were compared. The segmentation results of different methods on the SG-FCCM dataset were shown in Table \ref{Table1}, and five metrics, including Dice, mIoU, Precision, Recall, and F1-score, were used to evaluate the segmentation performance.

\begin{figure*}[h]
	\centering
	\includegraphics[scale=0.317]{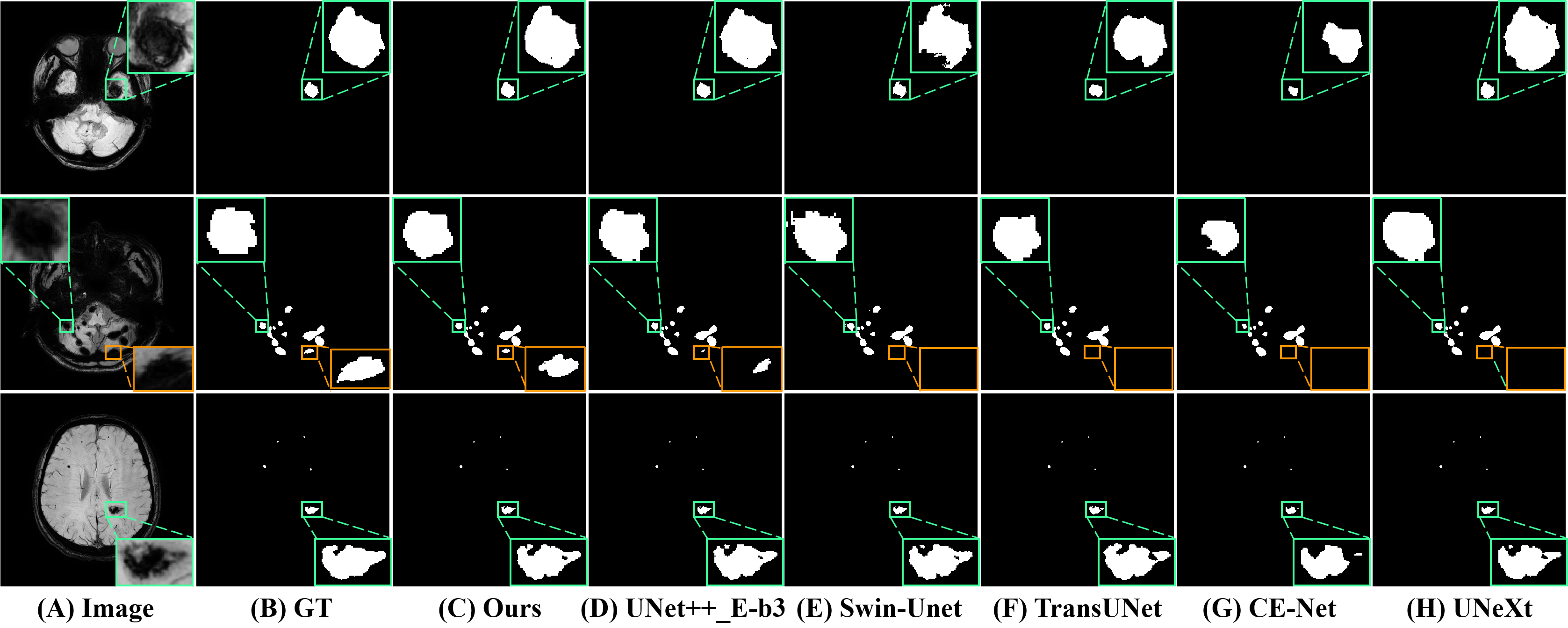}
	  \caption{Visualization results of different methods on the SG-FCCM dataset. Ours and UNet++\_E-b3 represent the model visualization results of U-Net and UNet++ as the backbone, and EfficientNet-B3 as the encoder, respectively.}
    \label{6}
    \vspace{-0.5em}
\end{figure*}

\begin{figure*}[h]
	\centering
	\includegraphics[scale=0.227]{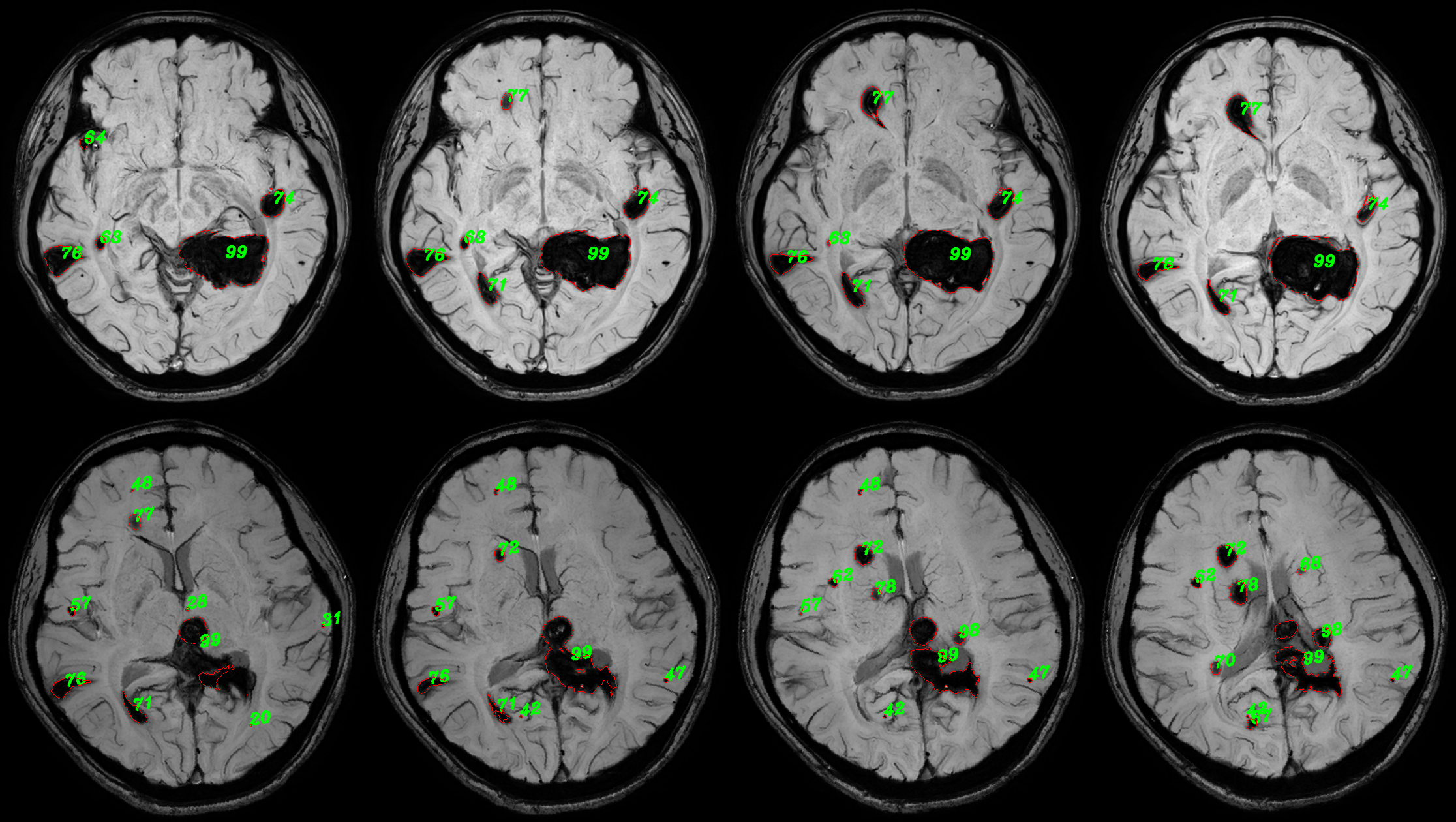}
	  \caption{Quantitative analysis visualization collage. This example consists of consecutive images from four planes of cranial MRI SWI sequences for two inspections of one patient. Numerals correspond to the indexes of the respective 3D lesion bodies. The red contour represents the visualization of the contours of the lesion segmentation predicted masks.}
    \label{7}
    \vspace{-0.5em}
\end{figure*}

It can be seen that the segmentation model with U-Net \cite{ronneberger2015u} architecture as the backbone and EfficientNet-B3 \cite{tan2019efficientnet} as the encoder achieved notable results in Dice (93.22\%), mIoU (90.76\%), Precision (92.19\%), and F1-score (86.45\%), demonstrating the best overall performance. The model achieved FLOPs score and parameter count of 48.48G and 2.55M, respectively. Despite the higher computational cost compared to the UNeXt \cite{valanarasu2022unext} model, it demonstrated superior segmentation performance. Therefore, this model was chosen as the network framework for FCCM lesion segmentation. Although the segmentation model using UNet++ \cite{zhou2018unet++} architecture as the backbone and EfficientNet-B3 \cite{tan2019efficientnet} as the encoder also achieved good performance, it was still slightly inferior to the model using U-Net \cite{ronneberger2015u} architecture as the backbone and incurred higher computational costs. Due to the limited volume of FCCM data and the presence of extremely small lesions in some images, networks such as DeepLabv3+ that utilized dilated convolutions, to a certain extent, reduced the network's sensitivity to local image details, making it challenging to capture small lesions effectively. This ultimately led to a slightly inferior segmentation performance. Visualization results of different methods were presented in Figure \ref{6}, affirming the capability of the chosen approach to accurately predict the location and boundaries of FCCM lesions.

\subsection{Quantitative statistics results}
In the FCCM lesion quantitative statistics module, 2D segmentation results of lesions were first acquired, and then a 3D segmentation mask was synthesized for quantitative statistics. The statistical information included the number and volume of lesions, and a comparative statistics of lesions between two examinations of the same patient was conducted. The quantitative statistical results of a case were presented to demonstrate the effectiveness of the proposed lesion quantitative statistical module. Figure \ref{7} showed the results of quantitative analysis visualization, where the numbering of each lesion and the contour of the lesion segmentation mask were plotted on the original FCCM image. The visualization collages of the entire volumes were shown in Figure \ref{10} and Figure \ref{11} in the Appendix A. The numberings and corresponding volume/volume pairs of lesions with a volume greater than 100 were presented in Table \ref{Table2}. The shaded areas depicted multiple lesions in the follow-up inspection that matched the same lesion in the previous inspection. This was mainly due to the irregular shape of the lesions, with discontinuities appearing in different planes. It manifested as a large lesion with several smaller lesions, resulting in a many-to-one situation. Figure \ref{7} could help doctors more intuitively compare the lesion and its segmentation mask, assist in distinguishing the lesion area, and combined with Table \ref{Table2} to conduct quantitative analysis on the lesion, making a more objective diagnosis of the condition.

\begin{table*}[]
\caption{\label{arttype2}Lesion number and corresponding volume information. The single volume and volume pair respectively represent the volume information of the lesion that is unsuccessfully matched and successfully matched in two examination results. The single volume denotes the volume of the lesion in the follow-up inspection, while the volume pair corresponds to the volumes of the lesion in the previous and follow-up inspections. }
\centering
\resizebox{\textwidth}{!}{%
\renewcommand{\arraystretch}{1.25}
\begin{tabular}{|cc|cc|cc|cc|cc|}
\hline
Lesion & Volume & Lesion & Volume         & Lesion & Volume             & Lesion & Volume                & Lesion & Volume                \\ \hline
...    & ...    & 51     & 168            & 59     & 639 & 67     & {[}1196, 1127{]} & 77     & {[}8108, 6903{]}   \\
44     & 103    & 52     & 196            & 60     & 1353   & 68     & {[}1379,   1331{]} & 78     & {[} 8479  ,   10042 {]} \\
45     & 108    & 53     & 196           & 61     & {[} 394 , 384 {]}   & 69     & {[}1662,   1687{]} & 94     & \cellcolor{lightgray!30}{[} 143305  ,   422 {]} \\
46     & 110    & 54     & 247 & 62     & {[} 571 , 543 {]}   & 70     & {[}2545, 2513{]}     & 95     & \cellcolor{lightgray!30}{[} 143305  ,   956 {]} \\
47     & 120    & 55     & 273 & 63     & {[} 735 , 689 {]}   & 71     & {[}2866,   2587{]}    & 96     & \cellcolor{lightgray!30}{[}143305 , 2747{]} \\
48     & 131    & 56     & 284 & 64     & {[}1063, 1310{]}   & 72     & {[}3387,   3446{]} & 97     & \cellcolor{lightgray!30}{[}143305 , 3783{]}  \\
49     & 132    & 57     & 353 & 65     & {[}1084, 1022{]}   & 74     & {[}6073, 5789{]}       & 98     & \cellcolor{lightgray!30}{[}143305 , 5857{]}   \\
50     & 157    & 58     & 411 & 66     & {[}1171, 1137{]}   & 76     & {[}7024, 6898{]}      & 99     & \cellcolor{lightgray!30}{[}143305, 23833{]}    \\ \hline
\end{tabular}
}
\label{Table2}
\vspace{-0.5em}
\end{table*}

\begin{figure*}[h]
	\centering
	\includegraphics[scale=0.065]{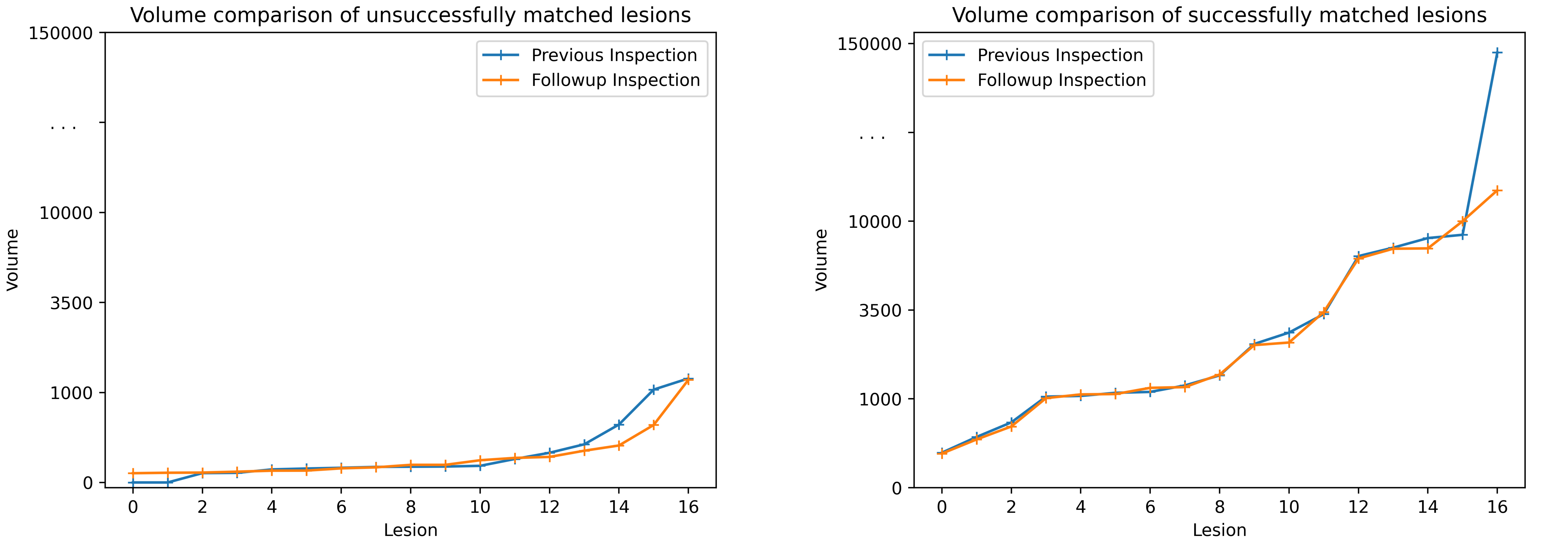}
	  \caption{Volume comparison of unsuccessfully matched and successfully matched lesions between the previous and follow-up inspection.}
    \label{8}
    \vspace{-0.5em}
\end{figure*}

\begin{table}[]
\caption{\label{arttype3}Classification statistics of lesion volume in follow-up inspection.}
\centering
\begin{tabular}{|c|c|}
\hline
Volume range & Number of lesions (Proportion) \\
\hline
$\leq$200 & 70 (71\%) \\
\hline
200-3500 & 22 (22\%) \\
\hline
$\geq$3500 & 7 (7\%) \\
\hline
\end{tabular}
\label{Table3}
\vspace{-0.5em}
\end{table}

In addition, further processing of quantitative statistical results could also help doctors have a better grasp of the patient's overall condition progression. For lesions with a volume greater than 100 in the two examination results, the volume broken line diagrams of unsuccessfully matched and successfully matched lesions were plotted in ascending order, as shown in Figure \ref{8}. By comparing the volume curves of lesions from two examinations, we could intuitively find that the volume of one lesion exhibited a significant reduction during the follow-up examination. Analysis in conjunction with Figure \ref{7} (see Figure \ref{10} and Figure \ref{11} in the Appendix A for details) indicated that this reduction was attributed to surgical intervention in the lesion area. No significant changes were observed in the overall size of other lesions. Simultaneously, a categorized statistical analysis of lesion volumes in the follow-up inspection was performed, as presented in Table \ref{Table3}. Based on these results, physicians could gain a clearer understanding of the quantity of severe brain lesions in patients than without using the quantitative statistical method, enabling better clinical decision-making.

\section{Discussion}
\subsection{Comparative analysis of annotation methods}
The scarcity of data has consistently posed a significant challenge in the application of deep learning in the medical field. In this study, to alleviate the annotation workload for doctors, a bounding box covering the target is used as a prompt, combined with BBTS and SAM for segmentation annotation. In the comparison of the segmentation annotation results of the two methods in Figure \ref{5}, the annotation of image B highlights the inadequacies of the SAM annotation method, mainly because the segmentation annotation of the SAM model depends on the performance of its own segmentation network. Recent studies \cite{GuochenNing20232023, diagnostics13111947} have indicated that due to a lack of medical image data and corresponding semantic masks, SAM's performance in zero-shot generalization on medical image datasets is not optimal, and thus performs poorly when dealing with regions with more complex and dense lesions. Image C illustrates the error-prone regions of the BBTS annotation method, primarily attributed to the similarity in pixel values between certain lesion areas and their surrounding regions, resulting in misannotations of non-lesion areas during threshold segmentation. From the figure, despite the inherent limitations of each approach, they are able to compensate for each other, thus obtaining more qualified annotation results than using only one segmentation annotation method.

\subsection{Efficiency of annotation strategy}
To validate the efficiency of the proposed annotation method, we select 50 FCCM images for annotation by clinicians, involving both contour tracing and bounding box annotation. Considering the feasibility of contour tracing annotation, images with moderate number and size of lesions are chosen. The time required for lesion contour annotation across the 50 images is approximately 96 minutes, while bounding box annotation takes only 25 minutes. Moreover, the screening of segmentation annotation results for SAM and BBTS takes less than 5 minutes. In summary, the annotation method we proposed can save doctors over 60\% of annotation time. Moreover, this has not taken into account the annotation of images with numerous and small lesions as shown in Figure \ref{1}. Subsequently, through iterative training annotation with segmentation networks, the workload for clinicians is further reduced, thereby enhancing annotation efficiency.

However, despite the reduction in the annotation time and associated workload for doctors achieved by our proposed annotation method, there are still challenges in obtaining qualified annotations for some lesion areas with complex target and background features. These regions are difficult to segment and obtain qualified annotations directly through bounding box prompts or iterative training. Manual annotation is still required to rectify the positions with disqualified masks, to which we will try to improve with better box-prompted segmentation methods or iterative training of segmentation models in our future work.

\subsection{Effects of quantitative statistical module}
Based on the experimental results, we analyze and discuss the effects of the quantitative statistical module. The statistical analysis of FCCM lesion volume and quantity plays a significant role in assisting clinical diagnosis and screening smaller lesions for early. However, currently widely adopted manual measurement methods often come with high time and resource costs. Additionally, different operators may interpret and standardize measurement results differently, leading to increased subjectivity and variability in measurements, inevitably introducing measurement errors. Therefore, a quantitative analysis of 3D masks synthesized based on FCCM segmentation results is performed. We quantify the volume by counting the number of pixels occupied by the lesions, calculate the number of lesions based on the segmented target regions, and establish a visualization framework. The quantitative analysis visualization collage presented in Figure \ref{7} (see Figure \ref{10} and Figure \ref{11} in Appendix A for details) enables doctors to quickly diagnose lesion areas and assess the accuracy of the predicted masks based on the original image. Lesion numbering assists doctors in locating corresponding lesion volume information, particularly the volume information for successfully matched lesions. Combining the overall trend of lesion volume changes between two examinations, as shown in Figure \ref{8}, provides better diagnostic insights into the patient's condition and disease progression and offers a quantitative assessment method for drug efficacy.

Moreover, in the application of assisting doctors in diagnosis, this module can set statistical thresholds as needed, ignoring lesions with volumes that are too small. This not only saves statistical time but also allows doctors to focus on important lesion areas. Future research will include discriminating the brain structures to which FCCM lesions belong and integrating our work into a system to present visualization and statistical results more effectively.

\section{Conclusion}
In this study, an FCCM lesion quantitative statistical framework is proposed, consisting of three components: efficient annotation module, FCCM lesion segmentation module, and FCCM lesion quantitative statistical module. We introduce a rapid and efficient data annotation method tailored for FCCM data, which combines binarization threshold segmentation annotation based on bounding boxes, SAM model segmentation annotation, and iterative training annotation of segmentation networks. The annotated data is utilized to train an image segmentation model, achieving high-precision segmentation of FCCM lesions. Finally, we present an FCCM lesion quantitative statistical module that, based on the segmentation predicted masks of FCCM lesions, computes the volume and quantity of lesions and realizes the quantitative comparison of lesions between multiple examinations of patients combined with image registration. In addition, a visualization system has been established to assist doctors in visually observing and quantitatively analyzing lesions. Experimental results on the SG-FCCM dataset demonstrate the effectiveness of our approach. This study addresses the limitations in existing FCCM lesion quantitative statistical methods, and we anticipate that the findings will contribute to the advancement of clinical diagnosis, treatment, and research in FCCM.

\section*{CRediT authorship contribution statement}
\textbf{Ruige Zong:}{ }Conceptualization, Methodology, Software, Validation, Data Curation, Writing - Original Draft, Writing - Review \& Editing, Formal analysis, Visualization. \textbf{Tao Wang:}{ }Conceptualization, Methodology, Formal analysis, Writing - Review \& Editing, Visualization. \textbf{Chunwang Li:}{ }Investigation, Data Curation, Resources, Writing - Review \& Editing. \textbf{Xinlin Zhang:}{ }Visualization, Writing - Review \& Editing. \textbf{Yuanbin Chen:}{ }Visualization, Writing - Review \& Editing. \textbf{Longxuan Zhao:}{ }Visualization. \textbf{Qixuan Li:}{ }Writing - Review \& Editing. \textbf{Qinquan Gao:}{ }Writing - Review \& Editing. \textbf{Dezhi Kang:}{ }Writing - Review \& Editing. \textbf{Fuxin Lin:}{ }Writing - Review \& Editing, Supervision, Project administration, Funding acquisition. \textbf{Tong Tong:}{ }Writing - Review \& Editing, Supervision, Project administration, Funding acquisition.

\section*{Acknowledgments}
\label{sec:Acknowledgments}
This work was supported by National Natural Science Foundation of China under Grant 62171133, the Fujian Health Commission under Grant 2022ZD01003, and Technology Platform Construction Project of Fujian Province under Grant 2020Y2003. 





\bibliographystyle{elsarticle-num}
\bibliography{sample}

\begin{thebibliography}{10}
\expandafter\ifx\csname url\endcsname\relax
  \def\url#1{\texttt{#1}}\fi
\expandafter\ifx\csname urlprefix\endcsname\relax\def\urlprefix{URL }\fi
\expandafter\ifx\csname href\endcsname\relax
  \def\href#1#2{#2} \def\path#1{#1}\fi

\bibitem{dalyai2011management}
R.~T. Dalyai, G.~Ghobrial, I.~Awad, S.~Tjoumakaris, L.~F. Gonzalez, A.~S. Dumont, N.~Chalouhi, C.~Randazzo, R.~Rosenwasser, P.~Jabbour, {Management of incidental cavernous malformations: a review}, Neurosurgical Focus FOC 31~(6) (2011) E5.
\newblock \href {https://doi.org/https://doi.org/10.3171/2011.9.FOCUS11211} {\path{doi:https://doi.org/10.3171/2011.9.FOCUS11211}}.

\bibitem{flemming2017population}
K.~D. Flemming, J.~Graff-Radford, J.~Aakre, K.~Kantarci, G.~Lanzino, J.~Brown, Robert~D., M.~M. Mielke, R.~O. Roberts, W.~Kremers, D.~S. Knopman, R.~C. Petersen, J.~Jack, Clifford~R., {Population-Based Prevalence of Cerebral Cavernous Malformations in Older Adults: Mayo Clinic Study of Aging}, JAMA Neurology 74~(7) (2017) 801--805.
\newblock \href {http://arxiv.org/abs/https://jamanetwork.com/journals/jamaneurology/articlepdf/2624328/jamaneurology\_flemming\_2017\_oi\_170012.pdf} {\path{arXiv:https://jamanetwork.com/journals/jamaneurology/articlepdf/2624328/jamaneurology\_flemming\_2017\_oi\_170012.pdf}}, \href {https://doi.org/10.1001/jamaneurol.2017.0439} {\path{doi:10.1001/jamaneurol.2017.0439}}.

\bibitem{zafar2019familial}
A.~Zafar, S.~A. Quadri, M.~Farooqui, A.~Ikram, M.~Robinson, B.~L. Hart, M.~C. Mabray, C.~Vigil, A.~T. Tang, M.~L. Kahn, H.~Yonas, M.~T. Lawton, H.~Kim, L.~Morrison, {Familial Cerebral Cavernous Malformations}, Stroke 50~(5) (2019) 1294--1301.
\newblock \href {http://arxiv.org/abs/https://www.ahajournals.org/doi/pdf/10.1161/STROKEAHA.118.022314} {\path{arXiv:https://www.ahajournals.org/doi/pdf/10.1161/STROKEAHA.118.022314}}, \href {https://doi.org/10.1161/STROKEAHA.118.022314} {\path{doi:10.1161/STROKEAHA.118.022314}}.

\bibitem{incerti2023magnetic}
I.~Incerti, M.~Fusco, V.~E. Contarino, S.~Siggillino, G.~Conte, S.~Lanfranconi, G.~A. Bertani, C.~Gaudino, P.~D~Orio, R.~Pallini, Q.~G. D~Alessandris, J.~M. T.~A. Meessen, E.~B. Nicolis, A.~Vasamì, E.~Dejana, A.~M. Bianchi, F.~M. Triulzi, R.~Latini, E.~Scola, {Magnetic susceptibility as a 1-year predictor of outcome in familial cerebral cavernous malformations: a pilot study}, European Radiology 33~(6) (2023) 4158--4166.
\newblock \href {https://doi.org/10.1007/s00330-022-09366-2} {\path{doi:10.1007/s00330-022-09366-2}}.

\bibitem{haasdijk2012cerebral}
R.~A. Haasdijk, C.~Cheng, A.~J. Maat-Kievit, H.~J. Duckers, {Cerebral cavernous malformations: from molecular pathogenesis to genetic counselling and clinical management}, European Journal of Human Genetics 20~(2) (2012) 134--140.
\newblock \href {https://doi.org/10.1038/ejhg.2011.155} {\path{doi:10.1038/ejhg.2011.155}}.

\bibitem{mouchtouris2015management}
N.~Mouchtouris, N.~Chalouhi, A.~Chitale, R.~M. Starke, S.~I. Tjoumakaris, R.~H. Rosenwasser, P.~M. Jabbour, S.~J. Monteith, {Management of Cerebral Cavernous Malformations: From Diagnosis to Treatment}, The Scientific World Journal 2015 (2015) 808314.
\newblock \href {https://doi.org/10.1155/2015/808314} {\path{doi:10.1155/2015/808314}}.

\bibitem{spiegler2018cerebral}
S.~Spiegler, M.~Rath, C.~Paperlein, U.~Felbor, {Cerebral Cavernous Malformations: An Update on Prevalence, Molecular Genetic Analyses, and Genetic Counselling}, Molecular Syndromology 9~(2) (2018) 60--69.
\newblock \href {http://arxiv.org/abs/https://karger.com/msy/article-pdf/9/2/60/3132779/000486292.pdf} {\path{arXiv:https://karger.com/msy/article-pdf/9/2/60/3132779/000486292.pdf}}, \href {https://doi.org/10.1159/000486292} {\path{doi:10.1159/000486292}}.

\bibitem{weinsheimer2023intracranial}
S.~Weinsheimer, J.~Nelson, A.~A. Abla, N.~U. Ko, C.~Tsang, O.~Okoye, J.~M. Zabramski, A.~Akers, A.~Zafar, M.~C. Mabray, B.~L. Hart, L.~Morrison, C.~E. McCulloch, H.~Kim, A.~A. Abla, A.~Akers, I.~A. Awad, B.~L. Hart, H.~Kim, N.~U. Ko, M.~T. Lawton, C.~Lee, M.~C. Mabray, C.~E. McCulloch, L.~Morrison, L.~Pawlikowska, E.~R. Smith, M.~Torbey, S.~Vadivelu, S.~Weinsheimer, J.~M. Zabramski, A.~Zafar, {Intracranial Hemorrhage Rate and Lesion Burden in Patients With Familial Cerebral Cavernous Malformation}, Journal of the American Heart Association 12~(3) (2023) e027572.
\newblock \href {http://arxiv.org/abs/https://www.ahajournals.org/doi/pdf/10.1161/JAHA.122.027572} {\path{arXiv:https://www.ahajournals.org/doi/pdf/10.1161/JAHA.122.027572}}, \href {https://doi.org/10.1161/JAHA.122.027572} {\path{doi:10.1161/JAHA.122.027572}}.

\bibitem{long2015fully}
J.~Long, E.~Shelhamer, T.~Darrell, {Fully Convolutional Networks for Semantic Segmentation}, in: Proceedings of the IEEE Conference on Computer Vision and Pattern Recognition (CVPR), 2015.

\bibitem{ronneberger2015u}
O.~Ronneberger, P.~Fischer, T.~Brox, {U-Net: Convolutional Networks for Biomedical Image Segmentation}, in: N.~Navab, J.~Hornegger, W.~M. Wells, A.~F. Frangi (Eds.), Medical Image Computing and Computer-Assisted Intervention -- MICCAI 2015, Springer International Publishing, Cham, 2015, pp. 234--241.

\bibitem{norman2018use}
B.~Norman, V.~Pedoia, S.~Majumdar, {Use of 2D U-Net Convolutional Neural Networks for Automated Cartilage and Meniscus Segmentation of Knee MR Imaging Data to Determine Relaxometry and Morphometry}, Radiology 288~(1) (2018) 177--185, pMID: 29584598.
\newblock \href {http://arxiv.org/abs/https://doi.org/10.1148/radiol.2018172322} {\path{arXiv:https://doi.org/10.1148/radiol.2018172322}}, \href {https://doi.org/10.1148/radiol.2018172322} {\path{doi:10.1148/radiol.2018172322}}.

\bibitem{sevastopolsky2017optic}
A.~Sevastopolsky, {Optic disc and cup segmentation methods for glaucoma detection with modification of U-Net convolutional neural network}, Pattern Recognition and Image Analysis 27~(3) (2017) 618--624.
\newblock \href {https://doi.org/10.1134/S1054661817030269} {\path{doi:10.1134/S1054661817030269}}.

\bibitem{roy2017relaynet}
A.~G. Roy, S.~Conjeti, S.~P.~K. Karri, D.~Sheet, A.~Katouzian, C.~Wachinger, N.~Navab, {ReLayNet: retinal layer and fluid segmentation of macular optical coherence tomography using fully convolutional networks}, Biomed. Opt. Express 8~(8) (2017) 3627--3642.
\newblock \href {https://doi.org/10.1364/BOE.8.003627} {\path{doi:10.1364/BOE.8.003627}}.

\bibitem{skourt2018lung}
B.~{Ait Skourt}, A.~{El Hassani}, A.~Majda, {Lung CT Image Segmentation Using Deep Neural Networks}, Procedia Computer Science 127 (2018) 109--113, pROCEEDINGS OF THE FIRST INTERNATIONAL CONFERENCE ON INTELLIGENT COMPUTING IN DATA SCIENCES, ICDS2017.
\newblock \href {https://doi.org/https://doi.org/10.1016/j.procs.2018.01.104} {\path{doi:https://doi.org/10.1016/j.procs.2018.01.104}}.

\bibitem{zhang2018road}
Z.~Zhang, Q.~Liu, Y.~Wang, {Road Extraction by Deep Residual U-Net}, IEEE Geoscience and Remote Sensing Letters 15~(5) (2018) 749--753.
\newblock \href {https://doi.org/10.1109/LGRS.2018.2802944} {\path{doi:10.1109/LGRS.2018.2802944}}.

\bibitem{apostolopoulos2017pathological}
S.~Apostolopoulos, S.~De~Zanet, C.~Ciller, S.~Wolf, R.~Sznitman, {Pathological OCT Retinal Layer Segmentation Using Branch Residual U-Shape Networks}, in: M.~Descoteaux, L.~Maier-Hein, A.~Franz, P.~Jannin, D.~L. Collins, S.~Duchesne (Eds.), Medical Image Computing and Computer Assisted Intervention - MICCAI 2017, Springer International Publishing, Cham, 2017, pp. 294--301.

\bibitem{alom2018recurrent}
M.~Z. Alom, M.~Hasan, C.~Yakopcic, T.~M. Taha, V.~K. Asari, {Recurrent Residual Convolutional Neural Network based on U-Net (R2U-Net) for Medical Image Segmentation}, arXiv preprint arXiv:1802.06955 (2018).

\bibitem{hoorali2022irunet}
F.~Hoorali, H.~Khosravi, B.~Moradi, {IRUNet for medical image segmentation}, Expert Systems with Applications 191 (2022) 116399.
\newblock \href {https://doi.org/https://doi.org/10.1016/j.eswa.2021.116399} {\path{doi:https://doi.org/10.1016/j.eswa.2021.116399}}.

\bibitem{flemming2020cerebral}
K.~D. Flemming, G.~Lanzino, {Cerebral Cavernous Malformation: What a Practicing Clinician Should Know}, Mayo Clinic Proceedings 95~(9) (2020) 2005--2020.
\newblock \href {https://doi.org/https://doi.org/10.1016/j.mayocp.2019.11.005} {\path{doi:https://doi.org/10.1016/j.mayocp.2019.11.005}}.

\bibitem{fox2021seizure}
C.~K. Fox, J.~Nelson, C.~E. McCulloch, S.~Weinsheimer, L.~Pawlikowska, B.~Hart, M.~C. Mabray, A.~Zafar, L.~Morrison, J.~M. Zabramski, A.~Akers, H.~Kim, {Seizure Incidence Rates in Children and Adults With Familial Cerebral Cavernous Malformations}, Neurology 97~(12) (2021) e1210--e1216.
\newblock \href {http://arxiv.org/abs/https://www.neurology.org/doi/pdf/10.1212/WNL.0000000000012569} {\path{arXiv:https://www.neurology.org/doi/pdf/10.1212/WNL.0000000000012569}}, \href {https://doi.org/10.1212/WNL.0000000000012569} {\path{doi:10.1212/WNL.0000000000012569}}.

\bibitem{cheng2021fetal}
D.~Cheng, X.~Shang, W.~Gao, F.~Barkhof, Y.~Liu, {Fetal Familial Cerebral Cavernous Malformation With a Novel Heterozygous KRIT1 Variation}, Neurology 97~(21) (2021) 986--988.
\newblock \href {http://arxiv.org/abs/https://www.neurology.org/doi/pdf/10.1212/WNL.0000000000012852} {\path{arXiv:https://www.neurology.org/doi/pdf/10.1212/WNL.0000000000012852}}, \href {https://doi.org/10.1212/WNL.0000000000012852} {\path{doi:10.1212/WNL.0000000000012852}}.

\bibitem{lan2023using}
J.~Lan, M.~Chen, J.~Wang, M.~Du, Z.~Wu, H.~Zhang, Y.~Xue, T.~Wang, L.~Chen, C.~Xu, et~al., {Using less annotation workload to establish a pathological auxiliary diagnosis system for gastric cancer}, Cell Reports Medicine 4~(4) (2023).

\bibitem{elaraby2021framework}
A.~Elaraby, I.~Elansary, {A Framework for Multi-Threshold Image Segmentation of Low Contrast Medical Images}, Traitement du Signal 38~(2) (2021) 309--314.

\bibitem{philbrick2019ril}
K.~A. Philbrick, A.~D. Weston, Z.~Akkus, T.~L. Kline, P.~Korfiatis, T.~Sakinis, P.~Kostandy, A.~Boonrod, A.~Zeinoddini, N.~Takahashi, B.~J. Erickson, {RIL-Contour: a Medical Imaging Dataset Annotation Tool for and with Deep Learning}, Journal of Digital Imaging 32~(4) (2019) 571--581.
\newblock \href {https://doi.org/10.1007/s10278-019-00232-0} {\path{doi:10.1007/s10278-019-00232-0}}.

\bibitem{deng2019new}
Y.~Deng, Y.~Sun, Y.~Zhu, Y.~Xu, Q.~Yang, S.~Zhang, Z.~Wang, J.~Sun, W.~Zhao, X.~Zhou, K.~Yuan, {A New Framework to Reduce Doctor's Workload for Medical Image Annotation}, IEEE Access 7 (2019) 107097--107104.
\newblock \href {https://doi.org/10.1109/ACCESS.2019.2917932} {\path{doi:10.1109/ACCESS.2019.2917932}}.

\bibitem{kline2016semiautomated}
T.~L. Kline, M.~E. Edwards, P.~Korfiatis, Z.~Akkus, V.~E. Torres, B.~J. Erickson, {Semiautomated Segmentation of Polycystic Kidneys in T2-Weighted MR Images}, AJR Am J Roentgenol 207~(3) (2016) 605.

\bibitem{racimora2010segmentation}
D.~Racimora, P.-H. Vivier, H.~Chandarana, H.~Rusinek, {Segmentation of polycystic kidneys from MR images}, in: N.~Karssemeijer, R.~M. Summers (Eds.), Medical Imaging 2010: Computer-Aided Diagnosis, Vol. 7624, International Society for Optics and Photonics, SPIE, 2010, p. 76241W.
\newblock \href {https://doi.org/10.1117/12.844361} {\path{doi:10.1117/12.844361}}.

\bibitem{yushkevich2006user}
P.~A. Yushkevich, J.~Piven, H.~C. Hazlett, R.~G. Smith, S.~Ho, J.~C. Gee, G.~Gerig, {User-guided 3D active contour segmentation of anatomical structures: Significantly improved efficiency and reliability}, NeuroImage 31~(3) (2006) 1116--1128.
\newblock \href {https://doi.org/https://doi.org/10.1016/j.neuroimage.2006.01.015} {\path{doi:https://doi.org/10.1016/j.neuroimage.2006.01.015}}.

\bibitem{sam2023segment}
A.~Kirillov, E.~Mintun, N.~Ravi, H.~Mao, C.~Rolland, L.~Gustafson, T.~Xiao, S.~Whitehead, A.~C. Berg, W.-Y. Lo, et~al., {Segment Anything}, arXiv preprint arXiv:2304.02643 (2023).

\bibitem{tan2019efficientnet}
M.~Tan, Q.~Le, {EfficientNet: Rethinking Model Scaling for Convolutional Neural Networks}, in: K.~Chaudhuri, R.~Salakhutdinov (Eds.), Proceedings of the 36th International Conference on Machine Learning, Vol.~97 of Proceedings of Machine Learning Research, PMLR, 2019, pp. 6105--6114.

\bibitem{10.1007/978-3-319-50835-1_22}
M.~A. Rahman, Y.~Wang, Optimizing intersection-over-union in deep neural networks for image segmentation, in: G.~Bebis, R.~Boyle, B.~Parvin, D.~Koracin, F.~Porikli, S.~Skaff, A.~Entezari, J.~Min, D.~Iwai, A.~Sadagic, C.~Scheidegger, T.~Isenberg (Eds.), Advances in Visual Computing, Springer International Publishing, Cham, 2016, pp. 234--244.

\bibitem{Rezatofighi_2019_CVPR}
H.~Rezatofighi, N.~Tsoi, J.~Gwak, A.~Sadeghian, I.~Reid, S.~Savarese, {Generalized Intersection Over Union: A Metric and a Loss for Bounding Box Regression}, in: Proceedings of the IEEE/CVF Conference on Computer Vision and Pattern Recognition (CVPR), 2019.

\bibitem{gu2019net}
Z.~Gu, J.~Cheng, H.~Fu, K.~Zhou, H.~Hao, Y.~Zhao, T.~Zhang, S.~Gao, J.~Liu, {CE-Net: Context Encoder Network for 2D Medical Image Segmentation}, IEEE Transactions on Medical Imaging 38~(10) (2019) 2281--2292.
\newblock \href {https://doi.org/10.1109/TMI.2019.2903562} {\path{doi:10.1109/TMI.2019.2903562}}.

\bibitem{cao2022swin}
H.~Cao, Y.~Wang, J.~Chen, D.~Jiang, X.~Zhang, Q.~Tian, M.~Wang, {Swin-Unet: Unet-Like Pure Transformer for Medical Image Segmentation}, in: L.~Karlinsky, T.~Michaeli, K.~Nishino (Eds.), Computer Vision -- ECCV 2022 Workshops, Springer Nature Switzerland, Cham, 2023, pp. 205--218.

\bibitem{zhang2021transfuse}
Y.~Zhang, H.~Liu, Q.~Hu, {TransFuse: Fusing Transformers and CNNs for Medical Image Segmentation}, in: M.~de~Bruijne, P.~C. Cattin, S.~Cotin, N.~Padoy, S.~Speidel, Y.~Zheng, C.~Essert (Eds.), Medical Image Computing and Computer Assisted Intervention -- MICCAI 2021, Springer International Publishing, Cham, 2021, pp. 14--24.

\bibitem{chen2021transunet}
J.~Chen, Y.~Lu, Q.~Yu, X.~Luo, E.~Adeli, Y.~Wang, L.~Lu, A.~L. Yuille, Y.~Zhou, {TransUNet: Transformers Make Strong Encoders for Medical Image Segmentation}, arXiv preprint arXiv:2102.04306 (2021).

\bibitem{valanarasu2022unext}
J.~M.~J. Valanarasu, V.~M. Patel, {UNeXt: MLP-Based Rapid Medical Image Segmentation Network}, in: L.~Wang, Q.~Dou, P.~T. Fletcher, S.~Speidel, S.~Li (Eds.), Medical Image Computing and Computer Assisted Intervention -- MICCAI 2022, Springer Nature Switzerland, Cham, 2022, pp. 23--33.

\bibitem{chen2018encoder}
L.-C. Chen, Y.~Zhu, G.~Papandreou, F.~Schroff, H.~Adam, {Encoder-Decoder with Atrous Separable Convolution for Semantic Image Segmentation}, in: Proceedings of the European Conference on Computer Vision (ECCV), 2018.

\bibitem{he2016deep}
K.~He, X.~Zhang, S.~Ren, J.~Sun, {Deep Residual Learning for Image Recognition}, in: Proceedings of the IEEE Conference on Computer Vision and Pattern Recognition (CVPR), 2016.

\bibitem{howard2017mobilenets}
A.~G. Howard, M.~Zhu, B.~Chen, D.~Kalenichenko, W.~Wang, T.~Weyand, M.~Andreetto, H.~Adam, {MobileNets: Efficient Convolutional Neural Networks for Mobile Vision Applications}, arXiv preprint arXiv:1704.04861 (2017).

\bibitem{zhou2018unet++}
Z.~Zhou, M.~M. Rahman~Siddiquee, N.~Tajbakhsh, J.~Liang, {UNet++: A Nested U-Net Architecture for Medical Image Segmentation}, in: D.~Stoyanov, Z.~Taylor, G.~Carneiro, T.~Syeda-Mahmood, A.~Martel, L.~Maier-Hein, J.~M.~R. Tavares, A.~Bradley, J.~P. Papa, V.~Belagiannis, J.~C. Nascimento, Z.~Lu, S.~Conjeti, M.~Moradi, H.~Greenspan, A.~Madabhushi (Eds.), Deep Learning in Medical Image Analysis and Multimodal Learning for Clinical Decision Support, Springer International Publishing, Cham, 2018, pp. 3--11.

\bibitem{deng2009imagenet}
J.~Deng, W.~Dong, R.~Socher, L.-J. Li, K.~Li, L.~Fei-Fei, {ImageNet: A large-scale hierarchical image database}, in: 2009 IEEE Conference on Computer Vision and Pattern Recognition, 2009, pp. 248--255.
\newblock \href {https://doi.org/10.1109/CVPR.2009.5206848} {\path{doi:10.1109/CVPR.2009.5206848}}.

\bibitem{GuochenNing20232023}
G.~Ning, H.~Liang, Z.~Jiang, H.~Zhang, H.~Liao, {The potential of 'Segment Anything' (SAM) for universal intelligent ultrasound image guidance}, BioScience Trends advpub (2023) 2023.01119.
\newblock \href {https://doi.org/10.5582/bst.2023.01119} {\path{doi:10.5582/bst.2023.01119}}.

\bibitem{diagnostics13111947}
P.~Shi, J.~Qiu, S.~M.~D. Abaxi, H.~Wei, F.~P.-W. Lo, W.~Yuan, {Generalist Vision Foundation Models for Medical Imaging: A Case Study of Segment Anything Model on Zero-Shot Medical Segmentation}, Diagnostics 13~(11) (2023).
\newblock \href {https://doi.org/10.3390/diagnostics13111947} {\path{doi:10.3390/diagnostics13111947}}.

\end{thebibliography}





\appendix
\section{Visualization supplement}
\setcounter{figure}{0}
\begin{figure*}[h]
	\centering
	\includegraphics[scale=0.11]{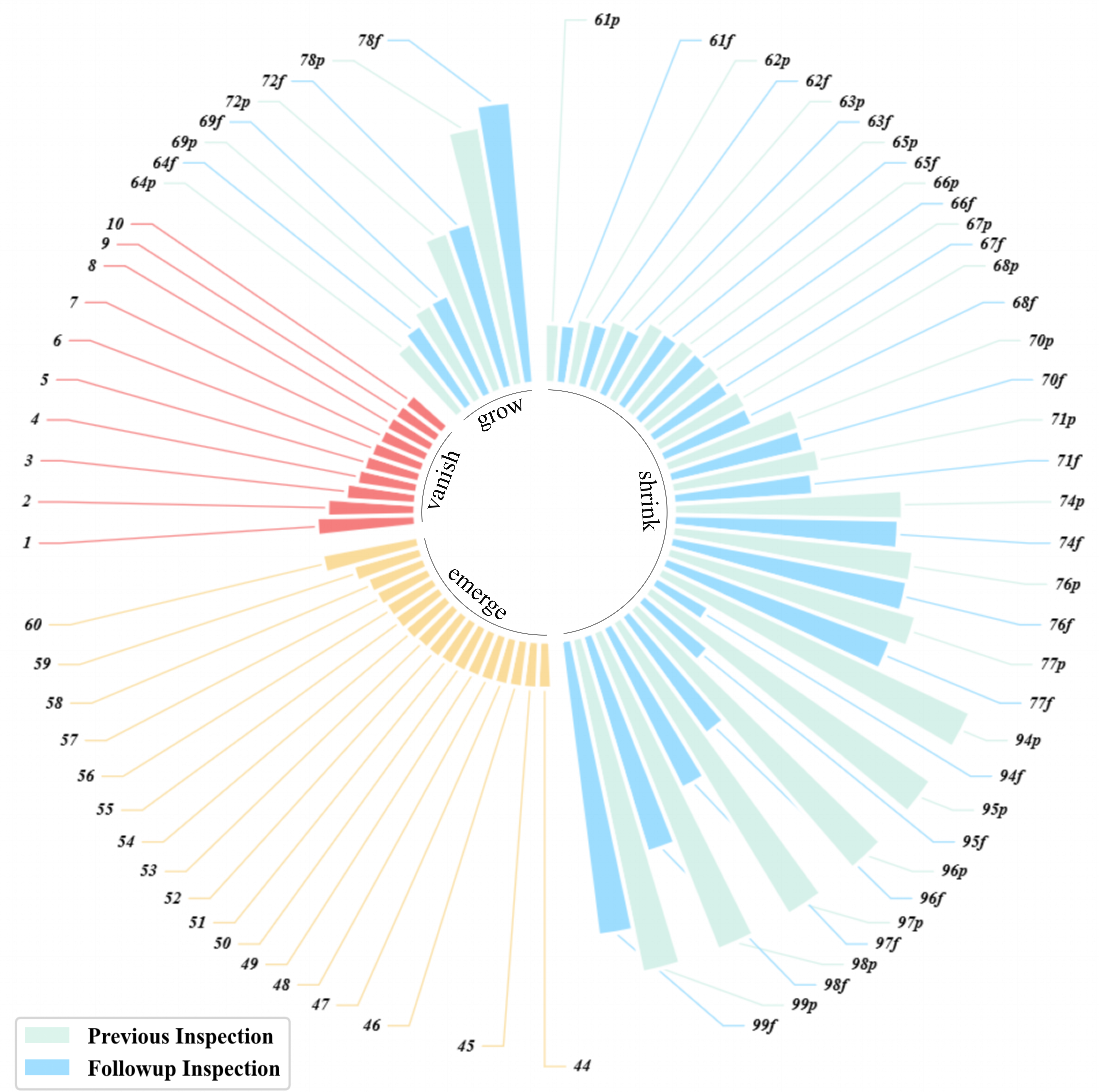}
	  \caption{FCCM quantitative statistical volume distribution.}
    \label{9}
    \vspace{-0.5em}
\end{figure*}

\begin{figure*}[h]
	\centering
	\includegraphics[scale=0.105]{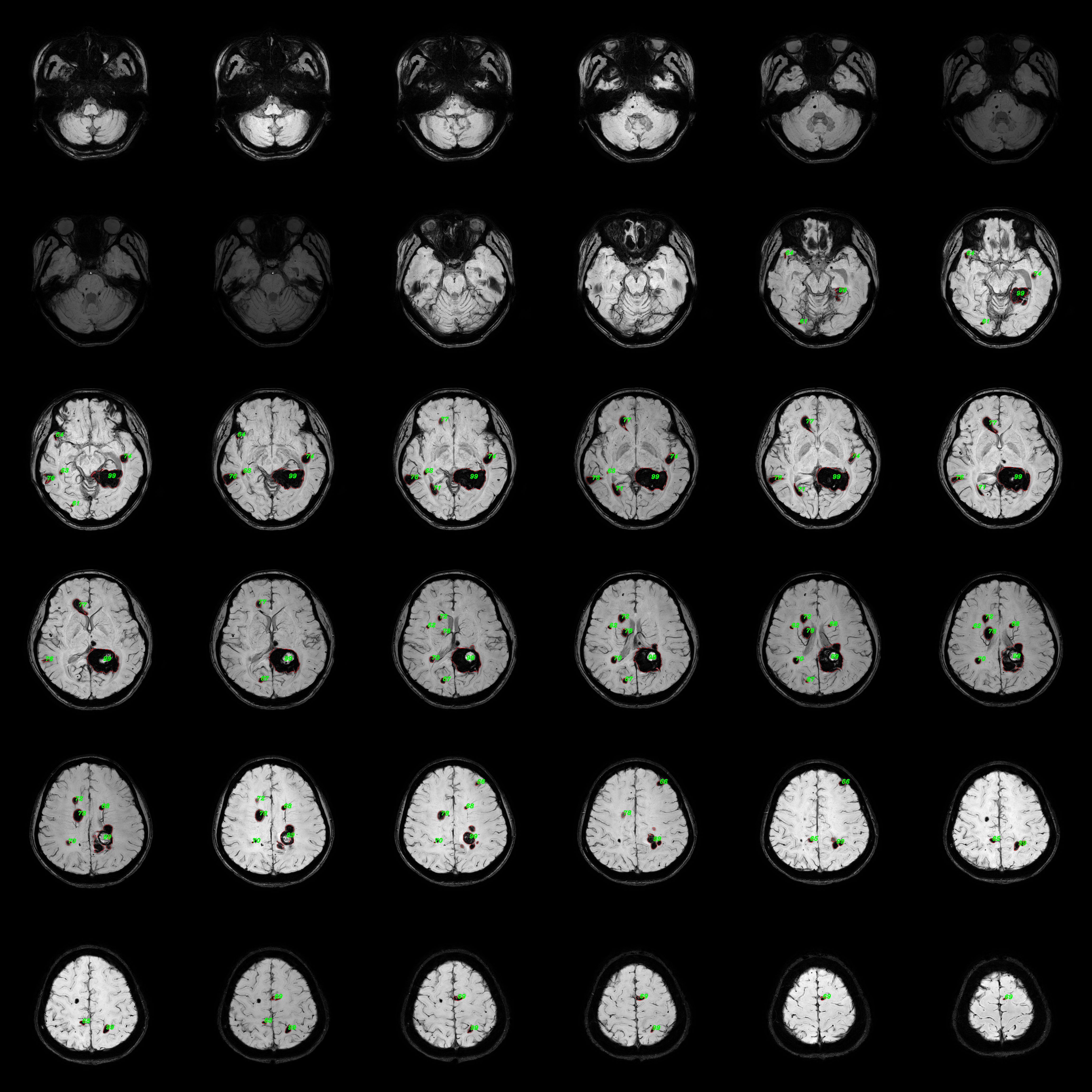}
	  \caption{Previous inspection visualization collage.}
    \label{10}
    \vspace{-0.5em}
\end{figure*}

\begin{figure*}[h]
	\centering
	\includegraphics[scale=0.105]{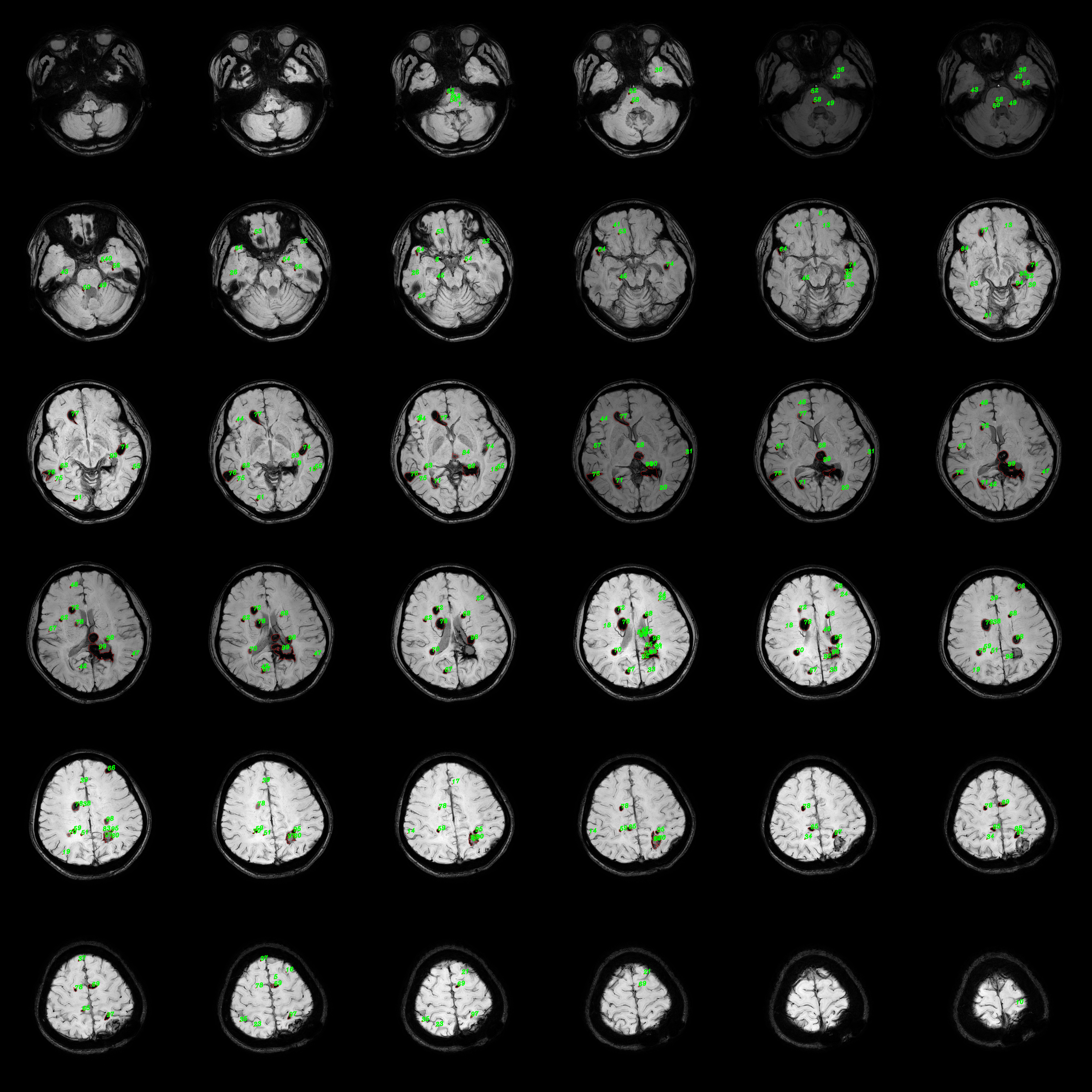}
	  \caption{Followup inspection visualization collage.}
    \label{11}
    \vspace{-0.5em}
\end{figure*}

\end{document}